\documentclass[twocolumn,showpacs,preprintnumbers,amsmath,amssymb,eqsecnum,pre]{revtex4}

\usepackage{graphicx}      
\usepackage{dcolumn}       

    \usepackage[usenames]{color}

\newcommand{\bm}{\mathbf}
\newcommand{\sm}{\mathsf}
\newcommand{\cm}{\mathcal}
\newcommand{\beq}{\begin{equation}}
\newcommand{\eeq}{\end{equation}}
\newcommand{\beqn}{\begin{eqnarray}}
\newcommand{\eeqn}{\end{eqnarray}}
\renewcommand{\Im}{\text{Im}}

\newcommand\beqa{\begin{eqnarray}}
\newcommand\eeqa{\end{eqnarray}}
\newcommand{\nn}{\nonumber\\}
\newcommand{\dd}{\text{d}}
\newcommand{\al}{\alpha}

\newcommand{\lx}{[\![}
\newcommand{\rx}{]\!]}
\newcommand{\NN}{\cm{N}}
\newcommand{\ii}{\text{i}}

\newcommand{\ed}{\end{document}}

\begin{document}
\title{Multicomponent fluids of hard hyperspheres in odd dimensions}

\author{Ren\'e D. Rohrmann}
\email{rohr@icate-conicet.gob.ar}
\homepage{http://icate-conicet.gob.ar/rohrmann}
\affiliation{Instituto de Ciencias Astron\'omicas, de la Tierra y del
Espacio (ICATE-CONICET), Avenida Espa\~na 1512 Sur, 5400 San Juan, Argentina}
\author{Andr\'es Santos}
\email{andres@unex.es}
\homepage{http://www.unex.es/eweb/fisteor/andres}
\affiliation{ Departamento de F\'{\i}sica, Universidad de Extremadura,
E-06071 Badajoz, Spain}

\date{\today}
\begin{abstract}

Mixtures of hard hyperspheres in odd space dimensionalities are studied with an analytical approximation method. This technique is based on the so-called rational function approximation and
provides a procedure for evaluating equations of state, structure factors, radial distribution functions, and direct correlations functions of additive mixtures of hard hyperspheres with any number of components and in arbitrary odd-dimension space. The method gives the exact solution of the Ornstein--Zernike equation coupled with the Percus--Yevick closure, thus extending  the solution  for hard-sphere mixtures [J. L. Lebowitz, Phys.\ Rev.\ \textbf{133}, 895 (1964)] to arbitrary odd
dimension.
Explicit evaluations for binary mixtures in five dimensions are performed. The results are compared with computer simulations and a good agreement is found.

\end{abstract}


\maketitle
\section{ Introduction} \label{s.intr}

Systems {made of hard bodies, i.e.,  impenetrable particles interacting solely through hard-core repulsions constitute useful simple fluid models}
 {\cite{M08}}.
Multicomponent hard-sphere mixtures serve as important reference systems
in condensed matter and are relevant for understating the behavior of
complex fluids with additional inter-particle interactions, such as those
in colloidal systems. These simple models capture the main features of the
packing effects at short distances as they occur also in fluids governed by
additional, attractive interactions. Besides, the knowledge of the structural
properties of hard sphere fluids is a prerequisite for treating attractive
interactions perturbatively and within a density functional theory approach.

The study of $d$-dimensional {hard-sphere} fluids may prove to be a useful guide
in investigating the solution of the three-dimensional (3D) problem, {apart from its own importance at a fundamental level.
This explains the continued interest on hard-hypersphere systems found in the literature along the years
\cite{RH64,FI81,LB82,J82,MT84,L84,BMS85,FRW85,FRW86,L86,KF86,CB86,FP87,R87,WRF87,BC87,R88,BR88,EF88,SMS89,ASV89,LM90,SM90,GGS90,MSAV91,GGS91,%
LZKH91,CFP91,GGS92,FP99,VMN99,BMC99,SYH99,MP99,PS00,YSH00,S00,YFP00,GAH01,FSL01,SYH01,SYH02,EAGB02,RHS04,BMV04,CM04,CM05,L05,SH05,L05b,BWK05%
,BMV05,BW05,CM06,LB06,SDST06,TS06,TS06b,TUS06,PZ06,RHS07,BW07,WBT07,RS07,BCW08,AKV08b,SST08,RRHS08,vMFC09,vMCFC09,LBW10,TS10,ER10}.}
However,  \emph{multicomponent} fluids at dimension $d>3$ have received much
less attention and the available information is rather sparse. The performed
studies reduce to some evaluations of equations of state and virial coefficients
\cite{R88,GGS92,SYH99,EAGB02}, phase transition analyses \cite{YSH00}, and
computer simulations \cite{GAH01}.
To our knowledge, {none} of the widely used mechanical-statistical theories {(e.g., integral equation theories)}
have been applied to these systems.

The Percus--Yevick (PY) theory \cite{PY58} is one of the {classical}
approximations of liquid-state theory and, certainly, one of the most widely used.
In the case of hard particles, exact PY solutions have been found for
single-component fluids in odd dimensions, $d=1$ \cite{ZP27}, $d=3$
\cite{W63, T63}, $d=5$ \cite{FI81,L84}, $d=7$ \cite{RHS04,RHS07}, $d=9,11$
\cite{RS07}, and also recently for even dimensions, $d=2$ \cite{AKV08} and
$d=4,6,8$ \cite{AKV08b}. Nevertheless, in the case of mixtures, PY solutions
were provided only for the 3D fluid \cite{L64, YSH98}, apart from
the exact solution known for the mixture of hard rods \cite{LPZ62}.
The present work is an attempt to cover the gap on hard particle mixtures at
dimensions higher than three using the so-called rational function
approximation (RFA).

The RFA approach was originally developed for hard-sphere fluids \cite{YS91}.
The method was successfully applied to {other related systems \cite{HYS08}, such as} hard-sphere mixtures
\cite{YSH98}, sticky hard spheres \cite{YS93, YS93b, SYH98}, square-well
fluids \cite{YS94, AS01}, penetrable spheres \cite{MYS07}, and one-component
hyperspheres \cite{RS07,RRHS08}. As we showed in Ref.\ \cite{RS07}, the RFA
method, in its simplest version, recovers the exact PY solution for
one-component hypersphere fluids in any space of odd dimension.

The aim of this paper is to extend the RFA theory to additive mixtures of
hyperspheres in odd-dimensional Euclidean space. {It is shown that the RFA method yields the exact solution of the
Ornstein--Zernike (OZ) equation with the PY closure}. While the method is generalized
to any odd dimension, we focus in particular on the solution for the five-dimensional (5D) system.. For this system, we  analyze some
of its thermodynamic and structural properties in the case of binary
mixtures, {and compare them with available simulation data}.

The remainder of this paper is organized as follows. Section \ref{s.ground} briefly
describes some basic quantities of the equilibrium theory for multicomponent
fluids of hard hyperspheres. In Sec.\ \ref{s.Gdef}, we introduce a Laplace
functional associated with the fluid structure factors and derive its general
properties. In Sec.\ \ref{s.rfa}, we present the extension of the  RFA method
{to} mixtures.
Explicit formulation for the 3D and 5D cases are given
in Sec.\ \ref{s.result}, {where compelling arguments about the equivalence to the PY solution are offered. Detailed evaluations for binary
mixtures in $d=5$ are presented in Sec.\ \ref{s.result5D}}. Section \ref{s.conclus} is devoted to the conclusions. {The most technical aspects of the paper are relegated to Appendices.}

\section{General background} \label{s.ground}

It is useful to give here some definitions that will be used in the following.
Let $\rho$ be the total number density of {an $\NN$-component}  hypersphere mixture, let $\{x_i\}$ (with $i=1,\ldots,\NN$) be the set
of mole fractions, and let $\{\sigma_i\}$ be the set of diameters. The overall packing fraction is $\eta=\sum_{i=1}^\NN\eta_i$,
where
\beq
\eta_i=v_d \rho x_i \sigma_i^d
\label{1}
\eeq
is the partial packing fraction due to species $i$. Here  $v_d$ is the volume of a $d$-dimensional
sphere of unit diameter.
For $d=\text{odd}$,
\beq
v_d=\frac{(\pi/2)^{(d-1)/2}}{d!!}.
\label{2}
\eeq

The structure factor $S_{ij}(k)$ for the particle pair ($i,j$) is given by
\beq \label{Sijh}
S_{ij}(k)=x_i \delta_{ij}+\rho x_i x_j \widehat{h}_{ij}(k),
\eeq
where
\beq
\widehat{h}_{ij}(k)=\int \dd \mathbf{r}\, h_{ij}(r)e^{-\ii \mathbf{k}\cdot\mathbf{r}}
\label{3}
\eeq
is the Fourier transform of the total pair
correlation function $h_{ij}(r)$, related to the radial distribution function
(rdf) $g_{ij}(r)$ by
\beq \label{hg}
h_{ij}(r)=g_{ij}(r)-1.
\eeq

The compressibility factor $Z$ may be written as
\beq \label{Zv}
Z\equiv \frac{p}{\rho k_BT}=1+\frac{2^{d-1}\eta}{\mu_d} \sum_{i,j=1}^\NN x_i x_j \sigma_{ij}^d g_{ij}
(\sigma_{ij}^+),
\eeq
where $p$ is the pressure, $k_B$ is the Boltzmann constant, $T$ is the temperature,
\beq
\mu_m \equiv  \sum_{\ell=1}^\NN x_\ell \sigma_\ell^m
\label{xi}
\eeq
denote the moments of the diameter distribution, and $g_{ij}(\sigma_{ij}^+)$
is the contact value of the rdf, $\sigma_{ij}$ being the minimum possible distance
 between particles $i$ and $j$.
For {\em additive} mixtures considered here one has
\beq
\sigma_{ij}=\frac{\sigma_i+\sigma_j}{2} .
\eeq
{In Eq.\ \eqref{Zv} we have used $\eta=v_d\rho\mu_d$.}

The isothermal susceptibility $\chi$ is given by
\beq \label{chi}
\chi^{-1}\equiv \frac{1}{k_BT}\left(\frac{\partial p}{\partial\rho}\right)_{T,\{x_i\}}=1-\rho \sum_{i,j=1}^\NN x_i x_j \widehat{c}_{ij}(0),
\eeq
where $\widehat{c}_{ij}(k)$ is the Fourier transform of the direct correlation
function $c_{ij}(r)$, which is defined by the OZ equation,
\beq \label{OZ}
\widehat{h}_{ij}(k)= \widehat{c}_{ij}(k)
+\rho \sum_{\ell=1}^\NN x_\ell \widehat{h}_{i\ell}(k) \widehat{c}_{\ell j}(k).
\eeq
In matrix form, the OZ relation can be rewritten as (see, e.g., Ref.\ \cite{YSH98})
\beq
\mathsf{I}-\widetilde{\mathsf{c}}(k)=\left[\mathsf{I}+\widetilde{\mathsf{h}}(k)\right]^{-1},
\label{4}
\eeq
where $\mathsf{I}$ is the $\NN\times \NN$ unit matrix and $\widetilde{\mathsf{c}}(k)$ and $\widetilde{\mathsf{h}}(k)$ are
$\NN\times \NN$ matrices  with elements $\rho \sqrt{x_i x_j} \widehat{c}_{ij}(k)$ and $\rho \sqrt{x_i x_j} \widehat{h}_{ij}(k)$, respectively.
The compressibility equation of state \eqref{chi} can be written as
\beqa
\chi^{-1}&=&
\sum_{i,j=1}^\NN \sqrt{x_i x_j}
\left[ \mathsf{I}-\widetilde{\mathsf{c}}(0)\right]_{ij}\nn
&=&\sum_{i,j=1}^\NN \sqrt{x_i x_j}\left[ \mathsf{I}+\widetilde{\mathsf{h}}(0)\right]^{-1}_{ij},
\label{chih}
\eeqa
where in the last step use has been made of  Eq.\ \eqref{4}.
In particular, in the case of binary mixtures ($\NN=2$), $\chi$ takes the form
\beq \label{chi_binary}
\chi=\frac{[1+\rho x_1 \widehat{h}_{11}(0)][1+\rho x_2 \widehat{h}_{22}(0)]
-\rho^2 x_1 x_2 \widehat{h}_{12}^2(0)}
{1+\rho x_1x_2 [\widehat{h}_{11}(0)+ \widehat{h}_{22}(0)-2 \widehat{h}_{12}(0)]}.
\eeq

The zero wavenumber value of $\widehat{h}_{ij}(k)$ can be expressed as
\beq \label{hij0}
\widehat{h}_{ij}(0)= d 2^d v_d H_{ij,d-1},
\eeq
where
\beq
H_{ij,m}= \int_0^\infty \dd r \, h_{ij}(r) r^{m}
\eeq
is the $m$th moment of $h_{ij}(r)$.
\section{The Laplace functional $G_{ij}(s)$} \label{s.Gdef}

\subsection{Definition}
In analogy to the case of one-component fluids \cite{RS07}, we introduce
the Laplace functional of the rdf in a Euclidean space of odd dimension $d$
as
\beq \label{Gdef}
G_{ij}(s)= \int_0^\infty \dd r\, r g_{ij}(r) \theta_n(sr) e^{-sr},
\eeq
which is defined in terms of the reverse Bessel polynomial $\theta_n(t)$ of
degree $n=(d-3)/2$:
\beq \label{theta}
\theta_n(t)= \sum_{\ell=0}^{n} \omega_{n,\ell} t^\ell,\quad \omega_{n,\ell}=
\frac{(2n-\ell)!}{2^{n-\ell}(n-\ell)!\ell!}.
\eeq
More details of these polynomials and their properties can be
found in Ref.\ \cite{RS07}. Here, we recall that  the
Fourier transform of the total correlation functions can be expressed in
terms of $G_{ij}(s)$ as (see Appendix \ref{a.fourier})
\beq \label{hij}
\widehat{h}_{ij}(k)= \nu_d
\left[ \frac{G_{ij}(s)-G_{ij}(-s)}{s^{d-2}}\right]_{s=\ii k},
\eeq
where
\beq
\nu_d\equiv (-2\pi)^{(d-1)/2}.
\eeq
The structure factors readily follow from Eq.\ \eqref{Sijh}.
We  {note} that the knowledge of $G_{ij}(s)$ allows us to obtain all the
structural and thermodynamic properties of a {multicomponent} hard $d$-sphere fluid.

\subsection{$G_{ij}(s)$ at long and short wavenumber} \label{s.Gls}

Being $g_{ij}(r<\sigma_{ij})=0$ for hard-hypersphere fluids, from Eq.\
(\ref{Gdef}), one obtains at long $s$ \cite{RS07}
\beq \label{Glong}
\lim_{s\rightarrow \infty} s^{(5-d)/2}\, e^{\sigma_{ij} s} G_{ij}(s)
=\sigma_{ij}^{(d-1)/2} \,g_{ij}(\sigma_{ij}^+).
\eeq
This equation determines the contact values of the radial distribution
function and we will use this to obtain the virial equation of the state of the
fluid through Eq.\ \eqref{Zv}.

On the other hand, from Eqs.\ (\ref{Gdef}) and (\ref{e.F1}), $G_{ij}(s)$ may be written
in terms of the total correlation function as
\beq \label{GsH}
G_{ij}(s)=\frac{(d-2)!!}{s^2}+
\int_0^\infty \dd r\, r h_{ij}(r) \theta_n(sr) e^{-sr}.
\eeq
The Taylor expansion of $e^{-sr}$ yields \cite{RS07}
\beq \label{Gshort}
G_{ij}(s)=\frac{(d-2)!!}{s^2} +\sum_{m=0}^\infty \alpha_{n,m} H_{ij,m+1} s^{m},
\eeq
where the numerical coefficients $\alpha_{n,m}$ are given by
\beq  \label{Ak}
\alpha_{n,m} = \sum_{\ell=0}^{\min(n,m)} \frac{(-1)^{m-\ell}}{(m-\ell)!}
\omega_{n,\ell}.
\eeq
The first $n$ coefficients $\alpha_{n,m}$ with $m=2q+1=\text{odd}$
($q=0,\ldots,n-1$) vanish \cite{RS07}. Therefore,
\beqa
 \label{GsHbis}
G_{ij}(s)&=&\frac{(d-2)!!}{s^2} +\sum_{m=0}^\infty \al_{n,2m} H_{ij,2m+1}
s^{2m}\nn &&+\sum_{m=n}^\infty \al_{n,2m+1} H_{ij,2m+2} s^{2m+1}.
\eeqa
As a consistency test, note that Eq.\ \eqref{hij0} is reobtained from Eqs.\ (\ref{hij})  and (\ref{GsHbis}), making use of Eq.\ \eqref{2} and $\alpha_{n,2n+1}=(-1)^{n+1}/(2n+1)!!$.
Therefore, the expansion of $G_{ij}(s)$ in powers of $s$ allows one to identify $H_{ij,d-1}$ from Eq.\ \eqref{GsHbis}. This in turn gives the compressibility equation of state via Eqs.\ \eqref{chih} and \eqref{hij0}.

In summary, the behaviors of $G_{ij}(s)$ at long and short $s$ are directly
connected to the thermodynamic variables $Z$ and $\chi$,
respectively.

\subsection{$G_{ij}(s)$ at low density} \label{s.Glrho}

The lowest terms in the density expansion of the rdf,
\beq
g_{ij}(r)= g_{ij}^{(0)}(r) + \rho g_{ij}^{(1)}(r) +O(\rho^2),
\eeq
are
\beq
g_{ij}^{(0)}(r)=\Theta(r-\sigma_{ij}),
\label{5}
\eeq
\beq
g_{ij}^{(1)}(r)=\Theta(r-\sigma_{ij}) \sum_\ell x_\ell
\Omega_{\sigma_{i\ell},\sigma_{j\ell}}(r),
\eeq
where  $\Theta(x)$ is Heaviside's step function and $\Omega_{a,b}(r)$ is the intersection volume of two
hyperspheres of radii ${a}$ and ${b}$ whose centers are
separated by a distance $r\leq a+b$. In Laplace space, one has
\beq \label{e.G01}
G_{ij}(s) = G_{ij}^{(0)}(s) + \rho G_{ij}^{(1)}(s) +O(\rho^2)
\eeq
with
\beq \label{e.G0}
G_{ij}^{(0)}(s)=\frac{\theta_{n+1}(\sigma_{ij}s)e^{-\sigma_{ij}s}}{s^2},
\eeq
\beq \label{e.G1}
G_{ij}^{(1)}(s) =  \sum_{\ell=1}^\NN x_\ell G_{ij\ell}^{(1)}(s),
\eeq
\beq \label{e.G1bis}
G_{ij\ell}^{(1)}(s) \equiv \int_{\sigma_{ij}}^\infty \dd r\,
r\theta_n(sr)\Omega_{\sigma_{i\ell},\sigma_{j\ell}}(r)e^{-sr} .
\eeq
%
\begin{figure}
\scalebox{0.42}{\includegraphics{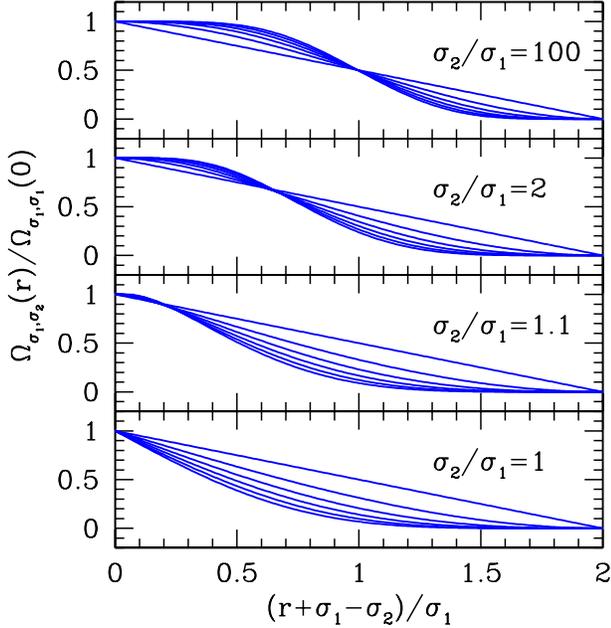}}
\caption{\label{f.overlap}
(Color online) Scaled overlap volume
$\Omega_{\sigma_1,\sigma_2}(r)/\Omega_{\sigma_1,\sigma_1}(0)$ of two
hyperspheres with radii $\sigma_1$ and $\sigma_2\geq \sigma_1$ as a function of the center
distance $r$, for Euclidean space with odd-dimension between 1 and 11 (lines
starting from the straight line), and for several values of the size ratio
$\sigma_2/\sigma_1$. The lower panel corresponds to two identical hyperspheres.
The limit $\sigma_2\rightarrow \infty$ with $\sigma_1$ fixed represents a
hypersphere of radius $\sigma_1$ crossing a flat wall.}
\end{figure}
In Eq.\ \eqref{e.G0} use has been made of Eq.\ \eqref{e.F2}. To obtain $G_{ij\ell}^{(1)}(s)$ we first need  the overlap volume $\Omega_{a,b}(r)$.
An explicit expression for the latter quantity, valid for arbitrary radii and $d=\text{odd}$, is derived in Appendix \ref{a.covolume} [see Eqs.\ \eqref{e.Omega} and \eqref{e.R4n4g}].
Figure \ref{f.overlap} shows graphs of the scaled intersection volume between
two hyperspheres of radii $\sigma_1$ and $\sigma_2\geq\sigma_1$ as a function of the
center separation $r$ (in rescaled units) for the first six odd-dimensions
($d=1,3,\ldots,11$), and for several size  ratios $\sigma_2/\sigma_1$.

{}Inserting Eq.\ \eqref{e.Omega} into the definition of $G_{ij\ell}^{(1)}(s)$ we get
\beq
 \label{e.Gijl}
G_{ij\ell}^{(1)}(s)
=(2\pi)^{(d-1)/2}\int_{\sigma_{ij}}^{\sigma_{i\ell}+\sigma_{j\ell}} \dd r\,
\frac{R_{4n+4}^{(\sigma_{i\ell},\sigma_{j\ell})}(r)}{r^{d-3}}  \theta_n(sr) e^{-sr},
\eeq
where we have taken into account that $\sigma_{ij}\geq |\sigma_{i\ell}-\sigma_{j\ell}|=|\sigma_i-\sigma_j|/2$.
It can be checked that $G_{ij\ell}^{(1)}(s)$ has the following structure:

\beq
G_{ij\ell}^{(1)}(s)=\frac{\nu_d}{s^{d-2}}\left[ G_{i\ell}^{(0)}(s)G_{j\ell}^{(0)}(s)+\frac{Q_{ij\ell}(s)}{s^{4}}e^{-\sigma_{ij}s}\right],
\label{6}
\eeq
where $Q_{ij\ell}(s)$ is a polynomial of degree  $3n+4=(3d-1)/2$ which can be further decomposed as
\beqa
Q_{ij\ell}(s)&=&s^{d+1} \overline{Q}_{ij\ell}(s)-\theta_{n+1}(\sigma_{i\ell}s)\sum_{m=0}^{n+1}\omega_{n+1,m}(\sigma_{j\ell} s)^m\nn
&&\times\sum_{q=0}^{2n+3-m}\frac{(-\sigma_\ell s)^q}{q!},
\label{3.1}
\eeqa
where $\overline{Q}_{ij\ell}(s)$ is a polynomial of degree $n=(d-3)/2$. Note that $Q_{ij\ell}(s)=Q_{ji\ell}(s)$ but $\overline{Q}_{ij\ell}(s)\neq \overline{Q}_{ji\ell}(s)$.

It is worth pointing out that all the equations in this section and in Sec.\ \ref{s.ground}
are exact.

\section{The Rational Function Approximation} \label{s.rfa}

In this section, we describe the RFA method to obtain the functional $G_{ij}(s)$,
which in turn allows us to obtain the structural and thermodynamic properties of the hypersphere mixture. The method presented here is, on  one hand, an
extension to mixtures of the one recently applied to one-component
systems of hyperspheres \cite{RS07} and, on the other hand, an extension to hyperspheres of the one already proposed for mixtures of hard spheres {($d=3$)} \cite{YSH98}.

The approximation we propose
consists of assuming the following functional form
\beq \label{Grfa}
G_{ij}(s) = \frac{e^{-\sigma_{ij}s}}{s^2}
\left[ \mathsf{L}(s)\cdot\mathsf{B}^{-1}(s)\right]_{ij},
\eeq
where $\mathsf{L}(s)$ and $\mathsf{B}(s)$ are $\NN\times \NN$ matrices  given by
\beq \label{Lserie}
\sm{L}(s)=\sum_{m=0}^{n+1} \sm{L}_m \, s^m,
\eeq
\beq \label{Bdef}
\sm{B}(s)= \sm{I}+\rho\sum_{m=0}^{n+1} \sm{\Phi}_m(s)\cdot\sm{L}_m.
\eeq
Here $\sm{L}_m$, $0\le m\le n+1$, are $\NN \times \NN$ matrices  whose elements
may depend on the fluid density, the particle diameters, and the mole fractions
of the system, but are independent of $s$. Therefore, $\sm{L}(s)$  has a polynomial dependence on $s$ of degree $n+1=(d-1)/2$.
Besides, $\sm{\Phi}_m(s)$ are {\emph{diagonal} matrices with elements} given by
\beq \label{phimatrix}
[\sm{\Phi}_m(s)]_{ii} = \nu_d       x_i \sigma_i^{d-m-\delta} s^{-\delta}
\phi_{d-m-\delta}(\sigma_i s),
\eeq
with
\beq  \label{phi}
\phi_m(x) \equiv \frac 1{x^{m}}
\left[ \sum_{\ell=0}^m \frac{(-x)^\ell}{\ell!} -e^{-x} \right].
\eeq
The series expansion  of $\sm{B}(s)$ and its asymptotic long-$s$ value are given in Appendix \ref{s.Bs}. The parameter $\delta$ in Eq.\ (\ref{phimatrix}) encompasses two different
conditions of normalization for $\sm{B}(s)$, specifically [see Eqs.\ \eqref{Bs0} and \eqref{Bsinfty}]
\beq \label{Ba}
\lim_{s\rightarrow 0} \sm{B}(s)=\sm{I} \quad (\delta=0), \quad
\lim_{s\rightarrow \infty} \sm{B}(s)=\sm{I} \quad (\delta=1).
\eeq
These cases constitute two alternative choices of the analytical
representation for $G_{ij}(s)$ which, however, yield identical physical
results (each one with its particular quantities $\sm{L}_m$).

In the {functional form \eqref{Grfa} for} $G_{ij}(s)$ we are using the constraints derived in
Sections \ref{s.Gls} and \ref{s.Glrho}.
In particular, as shown in Appendix \ref{s.lowdensity}, {Eq.\ (\ref{Grfa})} is consistent with the exact low-density expansion given by Eqs.\ (\ref{e.G01}), \eqref{e.G0}, \eqref{e.G1}, \eqref{6}, and \eqref{3.1}.

{The number $n+2$} of terms in the representations of $\sm{L}(s)$ and $\sm{B}(s)$,
Eqs.\ (\ref{Lserie}) and (\ref{Bdef}), is the {minimum one}  required to verify the correct
behavior of $G_{ij}(s)$ at large $s$. In fact, Eqs. (\ref{Glong}) and
(\ref{Grfa}) yield
\beqn \label{gc}
\sigma_{ij}^{n+1} g_{ij}(\sigma_{ij}^+) &=&
\lim_{s\rightarrow \infty} {s^{-(n+1)}}\left[\sm{L}(s)\cdot\sm{B}^{-1}(s)
\right]_{ij} \cr &=& \left[\sm{L}_{n+1}\cdot\sm{B}^{-1}(\infty)\right]_{ij},
\eeqn
with $\sm{B}(\infty)$ given by Eq.\ (\ref{Bsinfty}).
{With the normalization choice} $\delta=1$, the contact values $g_{ij}(\sigma_{ij}^+)$ are
directly related to the components of $\sm{L}_{n+1}$,
{
\beq \label{gc_1}
\sigma_{ij}^{(n+1)} g_{ij}(\sigma_{ij}^+) =
\left(
\sm{L}_{n+1}\right)_{ij}\quad (\delta=1).
\eeq
}

{Now we want to determine the $n+1$ coefficients {$\sm{L}_m$}. This is done by requiring consistency with Eq.\ \eqref{GsHbis}.  First, let us rewrite Eq.\ \eqref{Gshort} as
\beq
\frac{s^2}{(2n+1)!!}G_{ij}(s)=1+\sum_{m=0}^\infty \left(\sm{\Gamma}_m\right)_{ij} s^{m-2},
\eeq
where we have introduced the matrices $\sm{\Gamma}_m$ as
\beq
\label{defDG}
\left(\sm{\Gamma}_m\right)_{ij} \equiv \frac{ \alpha_{n,m}  H_{ij,m+1}}{(2n+1)!!}.
\eeq
Next, we note that
\beq
e^{\sigma_{ij}s}=\sum_{m=0}^\infty \left(\sm{\Lambda}_m\right)_{ij} s^m,
\eeq
where
\beq
\left(\sm{\Lambda}_m\right)_{ij} \equiv \frac{\sigma_{ij}^m}{m!}.
\eeq
Consequently,
\beq
\frac{s^2}{(2n+1)!!}e^{\sigma_{ij}s}G_{ij}(s)=\sum_{m=0}^\infty \left(\sm{K}_m\right)_{ij} s^{m},
\label{x1}
\eeq
}
with
\beq
\sm{K}_{m}\equiv \sm{\Lambda}_m +\sum_{\ell=0}^{m-2} \sm{\Gamma}_\ell \otimes
\sm{\Lambda}_{m-\ell-2},
\eeq
where the symbol
$\otimes$  denotes a matrix product ``element to element'':
$(A\otimes B)_{ij}\equiv A_{ij} B_{ij}$.

{Apart from the  {introduced notation}, Eq.\ \eqref{x1} is totally equivalent to Eq.\ \eqref{Gshort}. Now, according to the RFA form \eqref{Grfa},}
\beq \label{LKB}
\sum_{m=0}^{n+1} \widetilde{\sm{L}}_m \,s^m =
\left(\sum_{m=0}^\infty \sm{K}_m \, s^m \right) \cdot
\left(\sum_{k=0}^\infty \sm{B}_k \,s^k \right),
\eeq
where
\beq
\widetilde{\sm{L}}_m \equiv \frac{\sm{L}_m}{(2n+1)!!}
\eeq
{and the matrices $\sm{B}_k$ are given by Eqs.\ \eqref{B0} and \eqref{Bk}.}
From a power analysis of (\ref{LKB}), one obtains
\beq \label{KBL}
\sum_{k=0}^\ell \sm{K}_{\ell-k}\cdot\sm{B}_k=\widetilde{\sm{L}}_\ell,
\eeq
with the convention $\widetilde{\sm{L}}_\ell =0$ if $\ell>n+1$.
If we choose $\delta=0$, the first relation in (\ref{KBL})  (i.e., {$\ell=0$})
is trivially solved and yields
\beq \label{L0}
(\sm{L}_0)_{ij}=(2n+1)!!, \quad (\delta=0).
\eeq

Since, {as said below  Eq.\ \eqref{Ak},} the first $n$ coefficient $\alpha_{n,m}$ with $m=\text{odd}\leq 2n-1$
vanish, we have
\beq \label{gamma0}
\sm{\Gamma}_1 = \sm{\Gamma}_3 = \cdots = \sm{\Gamma}_{2n-1} = 0.
\eeq
The property (\ref{gamma0}) can be used, together with Eq.\ (\ref{KBL}) with
$\ell=\text{even}\leq 2n+2$, to express the matrices $\sm{\Gamma}_{2m}$,
$0\leq m\leq n$, in terms of the matrices $\sm{L}_\ell$ by means of the
recursion relation
\beq \label{gammare}
\sm{\Gamma}_{2m}=\widetilde{\sm{L}}_{2m+2}-\sum_{k=1}^{m+1} \sum_{\ell=0}^{2k}
\left[ \sm{\Gamma}_{2(m-k)}\otimes \sm{\Lambda}_{2k-\ell} \right] \cdot\sm{B}_l,
\eeq
where we must adopt {the convention} $\left(\sm{\Gamma}_{-2}\right)_{ij}\equiv 1$.
Similarly, from Eq. (\ref{KBL}) with $\ell=\text{odd}\leq 2n+1$ one obtains
\beqn \label{Lequ}
\widetilde{\sm{L}}_{2m+1}= \sum_{k=1}^{m+1} \sum_{\ell=0}^{2k-1}
\left[ \sm{\Gamma}_{2(m-k)}\otimes \sm{\Lambda}_{2k-\ell-1} \right] \cdot\sm{B}_\ell,
\eeqn
{ with $0\leq m\leq n$. Once the matrices $\sm{\Gamma}_{2m}$ are obtained in terms of $\sm{L}_\ell$ via Eq.\ \eqref{gammare}, Eq.\ \eqref{Lequ} becomes a closed system of $n+1=(d-1)/2$ matricial algebraic equations
for the unknowns $\sm{L}_1$, $\sm{L}_2$,\ldots , $\sm{L}_{n+1}$.}
Therefore, the problem of finding $G_{ij}(s)$ as given by Eq.\ (\ref{Grfa}) is
reduced to solving {Eqs.\ \eqref{gammare} and \eqref{Lequ}}.
In the case of hard-sphere mixtures {($d=3$)} it is possible to find an analytical
solution, as proved in Ref.\ \cite{YSH98}.

\begin{figure}
\scalebox{0.42}{\includegraphics{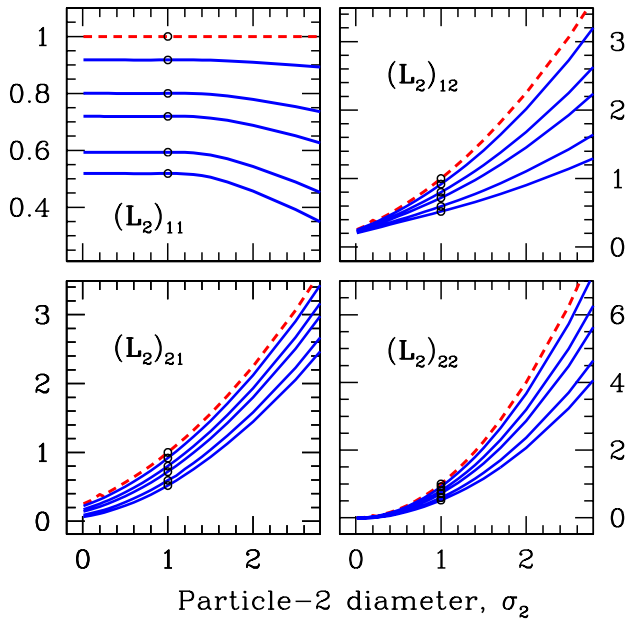}}
\caption{\label{f.matrix}
(Color online) Matrix element of $\sm{L}_2$ for a binary mixture at $d=5$ as functions
of the diameter $\sigma_2$ ($\sigma_1=1$) for {$x_2=\frac{1}{4}$} and different
values of the packing fraction. {}From top to bottom the curves correspond to
$\eta=0$, $0.01$, $0.02$, $0.05$, $0.1$ and $0.15$. The pure-fluid solutions
are indicated by symbols.}
\end{figure}
The solutions of Eq.\ (\ref{Lequ}) that are physically meaningful are those
that verify the correct behavior {in the limit $\rho\to 0$}. For $\delta=0$, we find from Eqs. (\ref{e.G01}), (\ref{e.G0}),
and (\ref{Grfa})
\beq \label{Lm0}
\lim_{\eta\to 0}\left(\sm{L}_m\right)_{ij}=\omega_{n+1,m}\sigma_{ij}^m,
\eeq
{where the coefficients $\omega_{n+1,m}$  of the reverse Bessel
polynomial $\theta_{n+1}(t)$ are given by Eq.\ \eqref{theta}.
Alternatively,}
when all particle diameters adopt the same value (say $\sigma_{ij}=1$ $\forall i,j$),
we must recover the pure-fluid solution, which only {depends} on density or
packing fraction (but not on {the mole fractions}), i.e.,
\beq \label{Lpf}
\left.(\sm{L}_m)_{ij}\right|_{\sigma_{ij}=1}=
a_m(\eta),
\eeq
{where the functions $a_m(\eta)$ are defined in Ref.\ \cite{RS07}.
Appendix \ref{s.one} shows that  the one-component solution \cite{RS07} is actually recovered
from the approach (\ref{Grfa}).}

Figure \ref{f.matrix} illustrates {the elements of the matrix $\sm{L}_2$, as obtained from  the physical solution of}
Eq.\ (\ref{Lequ}), for a 5D binary mixture  ($d=5$, $n=1$) as functions
of the diameter $\sigma_2$ (with $\sigma_1=1$) for {$x_2=\frac{1}{4}$}
and several density values. The dashed lines correspond to the limit of zero
density, as given by Eq.\ (\ref{Lm0}), while the symbols show the solution for
a one-component fluid [see Eq.\ (\ref{Lpf})], with $a_2(\eta)$ given by Eq.\ (E10) of Ref.\
\cite{RS07}.

Once the solutions of Eq.\ (\ref{Lequ}) are obtained, they may be used in
Eq.\ (\ref{gc}) to obtain the contact values of the pair {radial  distribution functions}
and, subsequently,  the compressibility factor in the so-called virial
route using the equation of state (\ref{Zv}). Its expression reads
\beq \label{Zvir}
Z_v=1+\frac{2^{d-1}\eta}{\mu_d} \sum_{i,j=1}^\NN x_i x_j {\sigma_{ij}^{(d+1)/2}}
\left[\sm{L}_{n+1}\cdot\sm{B}^{-1}(\infty)\right]_{ij}.
\eeq

Furthermore, the isothermal susceptibility $\chi$ of the fluid may be
evaluated by means of the thermodynamical formula (\ref{chi}), using the
relation resulting from Eq.\ (\ref{hij0}) and the definition of $\sm{\Gamma}_m$
in Eq.\ (\ref{defDG}),
\beq \label{hijg}
\widehat{\sm{h}}(0)= 2(-2\pi)^{(d-1)/2}(d-2)!! \sm{\Gamma}_{2n+1},
\eeq
with $\sm{\Gamma}_{2n+1}$ given by
\beq \label{Gammad2}
\sm{\Gamma}_{2n+1}= - \sum_{k=1}^{n+2} \sum_{\ell=0}^{2k-1}
\left[ \sm{\Gamma}_{2(n-k+1)}\otimes \sm{\Lambda}_{2k-\ell-1} \right] \cdot\sm{B}_l,
\eeq
which proceeds from Eq.\ (\ref{KBL}) at $\ell=2n+3$, {taking into account that $\widetilde{\sm{L}}_{2n+3}=\sm{0}$}. The values of the isothermal
susceptibility can be used then to obtain the compressibility factor
in the so-called compressibility route as
\beq  \label{Zcom}
Z_{c}(\eta)=\int_0^1 {\dd x}\,{\chi^{-1}(\eta x)}.
\eeq

Of course, the structure factors and total correlation functions can be
obtained easily with Eqs.\ (\ref{Sijh}) and (\ref{hij}), {once  the}
solutions $\sm{L}_m$ have been substituted into Eq.\ (\ref{Grfa}).
Notice that, {as shown in  Appendix \ref{s.lowdensity},} Eq.\ (\ref{Grfa})
is exact to first order in $\rho$, so that the first three terms of an
expansion of the right-hand side of Eq.\ (\ref{Sijh}) {in powers of
$\rho$ are exact within the RFA method.}

\section{Explicit expressions} \label{s.result}

In this section, we briefly revise the RFA solution for mixtures of hard
spheres ($d=3$, $n=0$) and derive results for 5D hyperspheres  ($d=5$, $n=1$).
The 3D analysis from the present framework is useful as a guide for
subsequent applications of the RFA method to mixtures in higher dimensions.
Here we use the version of the RFA approach based on the choice $\delta=0$.

For simplicity, henceforth, we introduce the following matrix notation
\beqn
\lx  A_{i} \rx_{\alpha\beta} &\equiv& A_{\alpha}, \quad
\lx  A_{j} \rx_{\alpha\beta} \equiv A_{\beta}, \cr
\lx  A_{k} \rx_{\alpha\beta} &\equiv& A_\alpha \delta_{\alpha \beta}, \quad
\lx  A_{ij} \rx_{\alpha\beta}\equiv A_{\alpha \beta}.
\label{notation}
\eeqn
{This means that, given a list $\mathcal{A}=\{A_1,A_2,\ldots,A_\NN\}$,  $\lx  A_{i} \rx$ represents the $\NN\times\NN$ square matrix made by repeating the list $\mathcal{A}$ as columns, so that  all the elements of a given row are equal. Analogously, $\lx  A_{j} \rx=\lx  A_{i} \rx^\dag$ (where $\dag$ indicates the transpose of a matrix) is obtained by repeating the list $\mathcal{A}$ as rows, so that  all the elements of a given column are equal. If the elements of $\mathcal{A}$ are placed along the main diagonal one gets the diagonal matrix $\lx  A_{k} \rx$. The meaning of $\lx  A_{ij} \rx$ is self-evident. We will also use the notation $\lx  1 \rx$  to refer to a matrix with all the elements equal to $1$.
Note the properties $\lx C_k\rx\cdot\lx A_j\rx=\lx C_iA_j\rx$, $\lx C_k\rx\cdot\lx A_i\rx=\lx C_iA_i\rx$, $\lx A_j\rx\cdot\lx C_k\rx=\lx C_jA_j\rx$,  $\lx A_i\rx\cdot\lx C_k\rx=\lx A_iC_j\rx$.
Some inversion properties involving these matrices are proved in Appendix \ref{s.matrix}.}

\subsection{Hard-sphere mixtures} \label{s.3D}

For $d=3$ ($n=0$), Eq. (\ref{Grfa}) becomes
\beq \label{Grfa3}
G_{ij}(s) = \frac{e^{-\sigma_{ij}s}}{s^2}
\left[ (\sm{L}_0+\sm{L}_1 s)\cdot \sm{B}^{-1}(s) \right]_{ij}
\eeq
with
\beq \label{d3_B}
\sm{B}(s)=\sm{I}+{\rho}\left[\sm{\Phi}_0(s)\cdot\sm{L}_0+\sm{\Phi}_1(s)\cdot\sm{L}_1\right],
\eeq
where $\sm{L}_0$ and $\sm{L}_1$ are the unknowns.
From Eq.\ (\ref{L0}) we have $\sm{L}_0=\lx 1 \rx$ and (\ref{Lequ}) yields
\beq \label{d3a_2}
\left(\sm{I} +\lx 2\eta_j  \rx \right)\cdot \sm{L}_1
=\lx \sigma_{ij}\rx +{\frac{\eta \mu_4}{2\mu_3}} \lx 1\rx,
\eeq
{with $\mu_m$ defined by Eq.\ \eqref{xi}.}
Using Eq.\ (\ref{invA}), it is straightforward to obtain
\beq \label{d3a_L1}
\sm{L}_1 =\frac {\lx \sigma_i\rx} 2  + \frac {\lx \sigma_j \rx}{2(1+2\eta)}
 -{\frac{\eta \mu_4 }{2\mu_3(1+2\eta)}} \lx 1 \rx.
\eeq
{This closes the determination of $G_{ij}(s)$.}

In order to evaluate the contact values of $g_{ij}$ [see Eq.\ (\ref{gc})],
\beq \label{d3gc}
\sigma_{ij} \, g_{ij}(\sigma_{ij}^+) =
\left[\sm{L}_1\cdot\sm{B}^{-1}(\infty)\right]_{ij},
\eeq
we must determine $\sm{B}(\infty)$. From Eq.\ (\ref{Bsinfty}) (with $\delta=0$),
{
\beq \label{B3da}
\sm{B}(\infty) = \sm{I}+\lx\Delta_i\rx+\lx\Omega_i\sigma_j\rx,
\eeq
with
\beq
\Delta_i \equiv
\frac{3\eta}{1+2\eta}\frac{\mu_4}{\mu_3}\frac{\eta_i}{\sigma_i}
- \eta_i  ,
\eeq
\beq
\Omega_i \equiv -\frac{3}{1+2\eta}  \frac{\eta_i}{\sigma_i}.
\eeq
}
Using the relation {(\ref{invABC})}, it results
{
\beqa \label{mez3d_Binv}
 \sm{B}^{-1}(\infty) =&=&
\sm{I}+\frac{1}{\gamma}\left\{ \alpha \lx \Omega_i\rx -(1+a)\lx \Omega_i \sigma_j\rx\right.\nn
&&\left.-(1+\beta)\lx \Delta_i\rx
+b\lx \Delta_i \sigma_j\rx \right\},
\eeqa
}
where
{
\beq \label{mez_3d_3}
a\equiv\text{tr}(\lx \Delta_i\rx)= -\eta\left(1+\beta\frac{\mu_4\mu_{2}}{\mu_3^2}\right),
\eeq
\beq
b\equiv\text{tr}(\lx \Omega_i\rx)=\beta\frac{\mu_{2}}{\mu_3},
\eeq
\beq
\alpha\equiv\text{tr}(\lx \Delta_i\sigma_j\rx)=-\frac{\eta\mu_4}{\mu_3}(1+\beta),
\eeq
\beq
\beta\equiv\text{tr}(\lx \Omega_i\sigma_j\rx)=-\frac{3\eta}{1+2\eta},
\eeq
\beq
\gamma\equiv(1+a)(1+\beta)-\alpha b=\frac{(1-\eta)^2}{1+2\eta}.
\eeq
The matrix products on the right-hand side of Eq.\ (\ref{d3gc}) are
\beqn \label{mez_3d_5}
\lx \sigma_i\rx\cdot\lx{\Delta_i}\rx &=& a \lx \sigma_i \rx, \quad
\lx \sigma_j\rx\cdot\lx{\Delta_i}\rx = \alpha \lx 1\rx, \nn
\lx \sigma_i\rx\cdot\lx{\Delta_i}\sigma_j\rx &=& a \lx \sigma_i\sigma_j\rx,\quad
\lx \sigma_j\rx\cdot\lx{\Delta_i}\sigma_j\rx = \alpha \lx \sigma_j\rx,\nn
\lx 1\rx\cdot\lx{\Delta_i}\rx &=& a \lx 1\rx
,\quad
\lx 1\rx\cdot\lx{\Delta_i}\sigma_j\rx = a \lx \sigma_j\rx,
\eeqn
plus similar results in the case of $\lx \Omega_i\rx$.
Taking into account the preceding results,
 one obtains from Eq.\ (\ref{d3gc})}
\beq \label{gd3_1}
g_{ij}(\sigma_{ij}^+) = \frac 1
{1-\eta}+{\frac{3\eta}{2(1-\eta)^2}\frac{\mu_2\sigma_i\sigma_j}{\mu_3\sigma_{ij}}}.
\eeq
Its application in Eq.\ (\ref{Zv}) gives the pressure equation in the virial route
obtained by Lebowitz \cite{L64}.

On the other hand, thanks to Eq.\  (\ref{d3a_L1}), {it is possible to solve Eqs.\ (\ref{hijg}) and
(\ref{Gammad2})  analytically}. The result is
{
\beqn \label{hd3}
\frac{\widehat{h}(0)}{\pi} &=& \frac{\eta}{8} \left(\frac{9\eta^2(\mu_4/\mu_3)^3}{(1+2\eta)^2}
-\frac{6\eta\mu_4\mu_5/\mu_3^2}{1+2\eta}+\frac{\mu_6}{\mu_3} \right) \lx 1 \rx  \cr
&& +\eta\frac{(1+2\eta)\mu_5/\mu_3-3\eta(\mu_4/\mu_3)^2}{(1+2\eta)^2}\lx \sigma_{ij} \rx
-\frac 18 \lx \sigma_i^3+\sigma_j^3\rx \cr
&& +\frac {\eta(\mu_4/\mu_3) (\lx \sigma_{ij}^2\rx + \lx\sigma_i\sigma_j\rx)
-\lx\sigma_i^2\sigma_j+\sigma_i\sigma_j^2\rx} {2(1+2\eta)}
\eeqn
}
The use of (\ref{hd3}) into (\ref{chi_binary}) yields the PY solution
for the isothermal susceptibility $\chi$, which agrees with that given by Ashcroft and
Langreth \cite{AL67} for a binary mixture of hard spheres.

\subsection{5D mixtures} \label{s.5D}

For a fluid of hyperspheres in $d=5$ ($n=1$), the functional $G_{ij}(s)$ takes
the form,
\beq \label{Grfa5}
G_{ij}(s) = \frac{e^{-\sigma_{ij}s}}{s^2}
\left[ (\sm{L}_0+\sm{L}_1 s+\sm{L}_2 s^2)\cdot\sm{B}^{-1}(s)\right]_{ij},
\eeq
with
\beq \label{Bsd5}
\sm{B}(s)= \sm{I}+{\rho}\left[\sm{\Phi}_0(s)\cdot\sm{L}_0+\sm{\Phi}_1(s)\cdot\sm{L}_1
+\sm{\Phi}_2(s)\cdot\sm{L}_2\right].
\eeq
According to Eq.\ (\ref{L0}), $\sm{L}_0=3 \lx1\rx$.
Besides, (\ref{Lequ}) at $m=0$ yields \cite{note}
\beq \label{d5_L1}
\sm{L}_1 = 3\lx \sigma_{ij}\rx + \frac{3}{1-6\eta} \left( 2{\eta\frac{\mu_6}{\mu_5}} \lx 1\rx +
3\eta \lx \sigma_j\rx - 10 \lx \eta_j/\sigma_j\rx \cdot\sm{L}_2 \right).
\eeq
{This expresses $\sm{L}_1$ in terms of $\sm{L}_2$.}
From Eq.\ (\ref{gammare}), with $m=0$, we find
{
\beqn \label{d5_gamma0}
\sm{\Gamma}_0 &=&
-\eta\left(\frac{9\mu_7/\mu_5}{14}+\frac{4\eta(\mu_6/\mu_5)^2}{1-6\eta}\right)
\lx 1\rx  - \frac {2\eta(\mu_6/\mu_5)  \lx \sigma_{ij}\rx }{1-6\eta}
  \cr
&& -\frac {3\eta \lx \sigma_i\sigma_j \rx} {2(1-6\eta)}- \frac 12 \lx \sigma_{ij}\rx^2 + \sm{Q}_{2} \cdot \sm{L}_{2},
\eeqn
where
\beq
\sm{Q}_{2} \equiv \frac{\sm{I}}{3}+ 3\lx \eta_j\rx
+ \frac {20\eta \mu_6 \lx \eta_j/\sigma_j\rx }{\mu_5(1-6\eta)}
+ \frac {5 \lx \eta_j \sigma_i/\sigma_j\rx }{1-6\eta}.
\eeq
}
Finally, using Eq.\ (\ref{Lequ}) with $m=1$, and the relations (\ref{d5_L1}) and
(\ref{d5_gamma0}), we obtain the following  quadratic equation for $\sm{L}_2$:
\beq \label{d5_L2}
\mathsf{0}= \sm{Q}_0 + \sm{Q}_1\cdot\sm{L}_2+\sm{Q}_{2}\cdot\sm{L}_{2}\cdot
\left( \sm{P}_{0}+\sm{P}_{1}\cdot\sm{L}_2 \right),
\eeq
where
\beq \label{d5_QP}
\sm{P}_{0} \equiv \frac 12\lx \sigma_k\rx +
2\lx \eta_i\sigma_i\rx +\frac{3 \lx\eta_i\sigma_j\rx}
{1-6\eta} + {\frac{12 \eta\mu_6 \lx\eta_i\rx}{\mu_5(1-6\eta)}},
\eeq
\beq
\sm{P}_{1} \equiv-\frac {60}{1-6\eta} \lx \eta_i\eta_j/\sigma_j\rx
 -10 \lx \eta_k/\sigma_k\rx,
\eeq
{
\beqa
\sm{Q}_0 &\equiv&- \eta\left[ \frac {20\eta^2(\mu_6/\mu_5)^3}{(1-6\eta)^2}+\frac{4\eta\mu_6\mu_7/\mu_5^2}{1-6\eta}
+\frac {\mu_8}{8\mu_5} \right] \lx 1 \rx \cr
&&- \eta\frac {10\eta(\mu_6/\mu_5)^2+(1-6\eta)\mu_7/\mu_5}{(1-6\eta)^2}\lx \sigma_{ij}\rx
 \cr
&&- \frac {2\eta\mu_6}{\mu_5(1-6\eta)}\lx \sigma_{ij}^2\rx
- \frac {3\eta}{2(1-6\eta)}\lx \sigma_i\sigma_j\sigma_{ij}\rx\nn
&&- \eta\frac {(1+24\eta)\mu_6}{4\mu_5(1-6\eta)^2}\lx \sigma_i\sigma_j\rx
-\frac  {\lx \sigma_{ij}^3 \rx} 3,
\eeqa
\beqa
\sm{Q}_1 &\equiv& \frac {10\eta\mu_6/\mu_5}{1-6\eta}\lx \eta_j \rx +\frac{\lx \sigma_k \rx}6
+\frac {5\lx \sigma_i \eta_j \rx }{2(1-6\eta)}+\frac 23 {\lx\eta_j\sigma_j \rx}\cr
&&
+ \eta\frac {100\eta(\mu_6/\mu_5)^2+10(1-6\eta)\mu_7/\mu_5}{(1-6\eta)^2}\lx \eta_j/\sigma_j\rx \cr
&& + \frac {25\eta\mu_6/\mu_5}{(1-6\eta)^2}\lx \sigma_i\eta_j/\sigma_j\rx
+ \frac {5 \lx \sigma_i^2\eta_j/\sigma_j\rx }{2(1-6\eta)}.
\eeqa
}

Due to the fact that Eq. (\ref{d5_L2}) is quadratic in $\sm{L}_2$ the
evaluation of the structure functions is now considerably more complex than
in the three-dimensional case. {In general,} solutions of (\ref{d5_L2}) must be worked {out}
numerically.

Binary systems can be completely specified
by the total packing fraction $\eta$, the concentration of one component (say $x_2$), and
the diameter ratio $\sigma_2/\sigma_1$. For arbitrary and finite values of
these parameters, Eq.\ (\ref{d5_L2}) yields four solutions $\sm{L}_2$ {that can be obtained analytically}, one
of which being the physical root that verifies the convergence conditions
(\ref{Lm0}) and (\ref{Lpf}).

\subsection{Direct correlation functions\label{s.cij}}
{The knowledge of the Laplace functions $G_{ij}(s)$ allows one to obtain the direct correlation functions via Eqs.\  \eqref{4} and \eqref{hij}. Although we have not attempted a formal proof, we have checked that in the binary case the structural form of $\widehat{c}_{ij}(k)$ is
\beqa
\widehat{c}_{ij}(k)&=&\frac{1}{k^{2d}}\left[\mathcal{P}_{ij}(\ii k)e^{\ii\sigma_{ij}k}+\mathcal{P}_{ij}(-\ii k)e^{-\ii\sigma_{ij}k}\right.\nn
&&\left.+\mathcal{Q}_{ij}(\ii k)e^{\ii(\sigma_i-\sigma_{j})k/2}+\mathcal{Q}_{ij}(-\ii k)e^{\ii(\sigma_j-\sigma_{i})k/2}\right],\nn
\label{cijk}
\eeqa
where $\mathcal{P}_{ij}(s)$ and $\mathcal{Q}_{ij}(s)$ are polynomials of
degrees $(3d-1)/2$ and $d-1$,
respectively. Moreover, the quantity
enclosed by square brackets is of order $k^{2d}$, so that
$\widehat{c}_{ij}(0)$ is finite, as required by Eq.\ \eqref{chih} and enforced through Eq.\ \eqref{GsHbis}.
}
{The direct correlation functions in real space, $c_{ij}(r)$ are obtained from Eq.\ \eqref{cijk} by means of Eq.\ \eqref{e.Fou1b}. The important point is that, upon application of the residue theorem, the structure given by Eq.\ \eqref{cijk} implies that $c_{ij}(r)=0$ for $r>\sigma_{ij}$ \cite{RS07}. Since the  RFA also complies with the physical
requirement $g_{ij}(r)=0$ for $r<\sigma_{ij}$, we recover the two conditions defining
precisely the PY closure to solve the OZ equation. Therefore, we find that the  RFA developed in this paper yields the PY solution for hard-hypersphere mixtures
of odd dimensions. This is a remarkable result
since both approaches (RFA and PY) are, in principle, rather independent.}

\begin{figure}
\scalebox{0.42}{\includegraphics{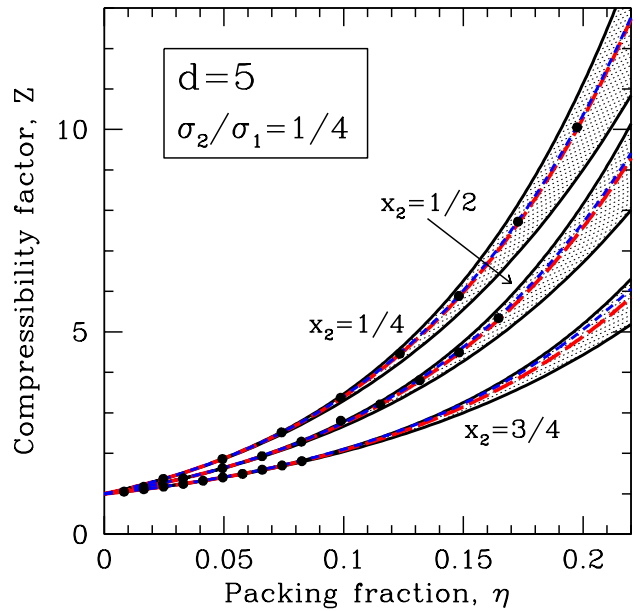}}
\caption{\label{f.Z5eta} (Color online) Compressibility factor as a function of the packing
fraction $\eta$ for a binary mixture at $d=5$ with $\sigma_2/\sigma_1=\frac{1}{4}$
and  {mole fractions} $x_2=\frac{1}{4}$, $\frac{1}{2}$ and $\frac{3}{4}$ (indicated on the plot).
{RFA-PY} results obtained from the virial and the compressibility routes (solid
lines on the right and left of the shaded area, respectively) are compared
with predictions of an analytical equation proposed in Ref.\ \protect\cite{SYH99} ({short dashed}
lines) and simulation results from Ref.\ \protect\cite{GAH01} (symbols). {The {long dashed} lines represent the interpolation $Z=\frac{2}{5}Z_v+\frac{3}{5}Z_c$.}}
\end{figure}

\section{Results for 5D binary mixtures} \label{s.result5D}

Here we consider a  mixture of two types ($i=1,2$) of 5D spheres
with arbitrary diameters $\sigma_i$ and concentrations $x_i$.
For simplicity we fix $\sigma_1=1$.

Figure \ref{f.Z5eta} compares the compressibility factors calculated from
the virial ($Z_v$) and compressibility ($Z_c$) routes [cf.\ Eqs.\ (\ref{Zvir}) and
(\ref{Zcom}), respectively] with those from computer simulations (molecular
dynamics) \cite{GAH01} and a {semiempirical} equation of state \cite{SYH99},
for a mixture with a diameter ratio $\sigma_2/\sigma_1=\frac{1}4$ and {mole fractions}
$x_2=\frac{1}4$, $\frac{1}2$ and $\frac{3}4$.
{We observe} that $Z_c$ and $Z_v$ bound both the simulation data and the
{proposal of Ref.\ \cite{SYH99}}, with $Z_c$ slightly above and $Z_v$ slightly below.
The agreement is very good at low densities (packing fraction lower than $0.1$)
and reasonably good over the whole density range ($0\le \eta < 0.19$) of
fluid phase predicted for the one-component system. In general, the
compressibility route gives better results than the virial one.
A similar relation $Z_v < Z < Z_c$, with $Z$ being the actual compressibility factor,
has been observed for the PY solution of the pure 5D system \cite{RS07}.
{Figure \ref{f.Z5eta} also includes the interpolation formula $Z=\frac{2}{5}Z_v+\frac{3}{5}Z_c$ \cite{S00}. This Carnahan--Starling-like recipe presents a very good agreement with simulation data and is practically indistinguishable from the semi-empirical equation of state proposed in Ref.\ \cite{SYH99}, except for $x_2=\frac{3}{4}$ and $\eta>0.15$. Unfortunately, no simulation data are available in those cases.}

\begin{figure}
\scalebox{0.42}{\includegraphics{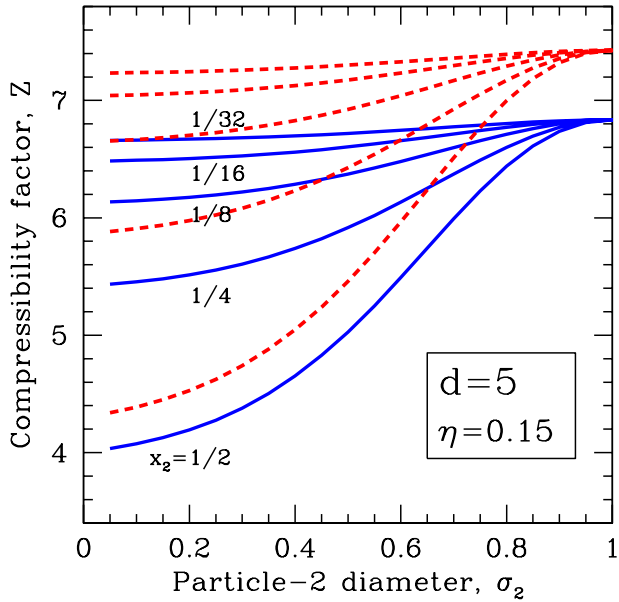}}
\caption{\label{f.Z5sig} (Color online) Compressibility factor obtained from the {RFA-PY} method
for a binary mixture at $d=5$, as a function of the particle diameter
$\sigma_2$ ($\sigma_1=1$) for $\eta=0.15$ and {mole fractions} $x_2$ indicated
on the plot (solid lines). Predictions of the analytical equation from
Ref.\ \protect\cite{SYH99} are shown with {dashed} lines.}
\end{figure}
Next, it is instructive to examine the size ratio dependence of the
compressibility factor. Figure \ref{f.Z5sig} compares values of $Z_v$ obtained
from the {RFA-PY solution}  via the virial route with those calculated from the analytical
expression given in Ref.\ \cite{SYH99}. These evaluations correspond to a packing
fraction $\eta=0.15$ and a set of molar fractions {ranging from $x_2=\frac{1}{32}$ to $x_2=\frac{1}{4}$}.
As the concentration $x_2$ {decreases}, we find, as expected, that the
compressibility factor comes to have a weaker dependence on $\sigma_2$.
The general trends of $Z_v$ versus $\sigma_2$ for different {mole fractions}
$x_2$ {qualitatively} agree with the predictions of the  equation of state proposed
in Ref.\ \cite{SYH99} (there are no simulation data available for this analysis).
The differences between both calculations (at most 9\%) are similar
to those observed before.

\begin{figure}
\scalebox{0.44}{\includegraphics{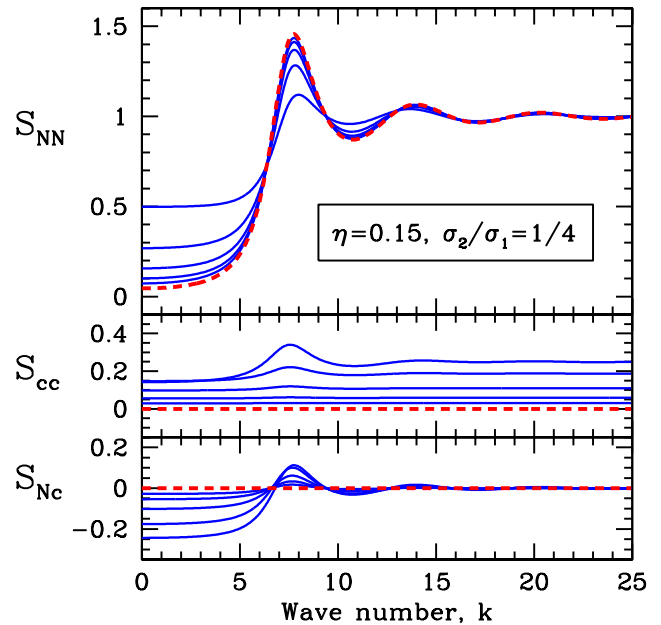}}
\caption{\label{f.SNN} (Color online) Number-concentration structure factors $S_{NN}$,
{$S_{Nc}$} and {$S_{cc}$}, as obtained from the {RFA-PY} method for a binary mixture in $d=5$ {dimensions}
with diameter ratio $\sigma_2/\sigma_1=\frac{1}{4}$, packing fraction $\eta=0.15$, and
{mole} fractions $x_2=\frac{1}2$, $\frac{1}4$,
$\frac{1}8$, $\frac{1}{16}$, and $\frac{1}{32}$ (solid lines).
As $x_2$ is reduced, the curves converge to the solution of the pure fluid
(dashed lines).}
\end{figure}

{Now we analyze the structure and correlation functions.} Instead {of  the conventional structure factors $S_{ij}$, we consider some combinations of them which may be easily associated with fluctuations of the thermodynamic variables \cite{AL67,BT70}:}
\begin{subequations}
\label{SNc}
\beq
S_{NN}(k)= S_{11}(k) + S_{22}(k) + 2 S_{12}(k),
\eeq
\beq
{S_{Nc}}(k)= x_2 S_{11}(k) - x_1 S_{22}(k) + (x_2-x_1) S_{12}(k),
\eeq
\beq
{S_{cc}}(k)= x_2^2 S_{11}(k) + x_1^2 S_{22}(k) - 2 x_1 x_2 S_{12}(k).
\eeq
\end{subequations}
In the limit of small wavenumber ($k\rightarrow0$), $S_{NN}$ and
{$S_{cc}$} become the mean square fluctuations in the particle number and
concentration, respectively, whereas {$S_{Nc}$} is the correlation between
these two fluctuations.

{We
plot in Fig.\ {\ref{f.SNN}} the number-concentration structure
factors for} a packing fraction $\eta=0.15$, a diameter ratio $\sigma_2/\sigma_1=\frac{1}{4}$,
and decreasing values of the particle-$2$ concentration, $x_2=\frac{1}2$, $\frac{1}4$,
$\frac{1}8$, $\frac{1}{16}$, and $\frac{1}{32}$. As  expected, {$S_{cc}$} and {$S_{Nc}$} reduce
smoothly to zero with $x_2$ for all wavenumber, whereas the number-number structure
factor $S_{NN}$ converges to the static structure factor of the one-component
fluid in this limit (dashed line).
Notice that $S_{NN}$ and {$S_{cc}$} are positive for all $k$ by definition.
The main modifications of the structure factor $S_{NN}$ with increasing $x_2$
 are the variation of height of the
{main} peak, with {its} position very weakly altered, and an {increase}
of the values at {short $k$}.

\begin{figure}
\scalebox{0.44}{\includegraphics{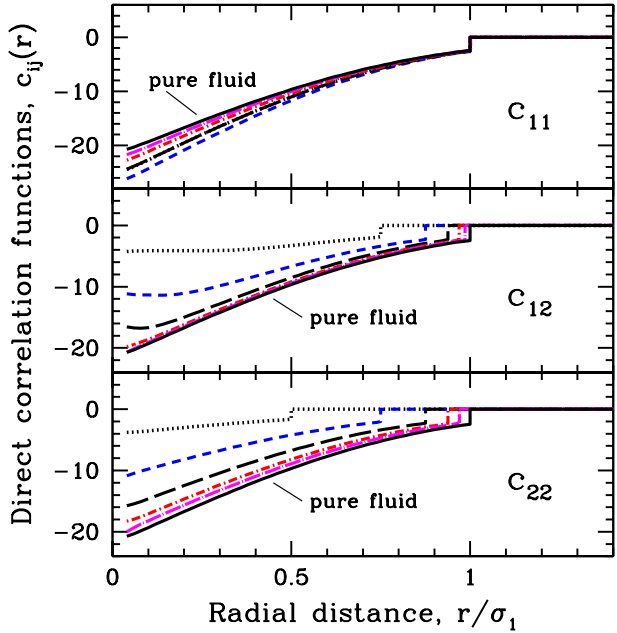}}
\caption{\label{f.cr_rho14} (Color online) Direct correlation functions obtained from the {RFA-PY}
method for an equimolar ($x_1=x_2=\frac{1}{2}$) binary mixture in $d=5$ {dimensions} at
$\eta=0.15$ and various diameter ratios,
$\sigma_2/\sigma_1=\frac{1}2$ (dotted line), $\frac{3}4$
(dashed line), $\frac{7}8$ (long dashed line), $\frac{15}{16}$ (dashed-dotted line), and
$\frac{31}{32}$ (long dashed-dotted line). The pure-fluid solution is displayed
for comparison (solid line).}
\end{figure}
Evaluations of the direct correlation functions defined (in the Fourier
space) by the OZ equation (\ref{OZ}) are shown in
Fig.\ \ref{f.cr_rho14}. The results correspond to an equimolar mixture for
different diameter ratios from $\sigma_2/\sigma_1=\frac{1}{2}$ to nearly
equivalent hypersphere sizes, $\sigma_2/\sigma_1=\frac{31}{32}$.
In the limit $\sigma_2/\sigma_1 \rightarrow 1$ the three correlation
functions $c_{ij}(r)$ become identical and match the pure-fluid values.
{Figure \ref{f.cr_rho14} clearly shows that the direct correlation functions
$c_{ij}(r)$ vanish for $r>\sigma_{ij}$, thus confirming  that the RFA method
yields solutions of the PY closure to the OZ
equation, as discussed in Sec.\ \ref{s.cij}}.

\begin{figure}
\scalebox{0.42}{\includegraphics{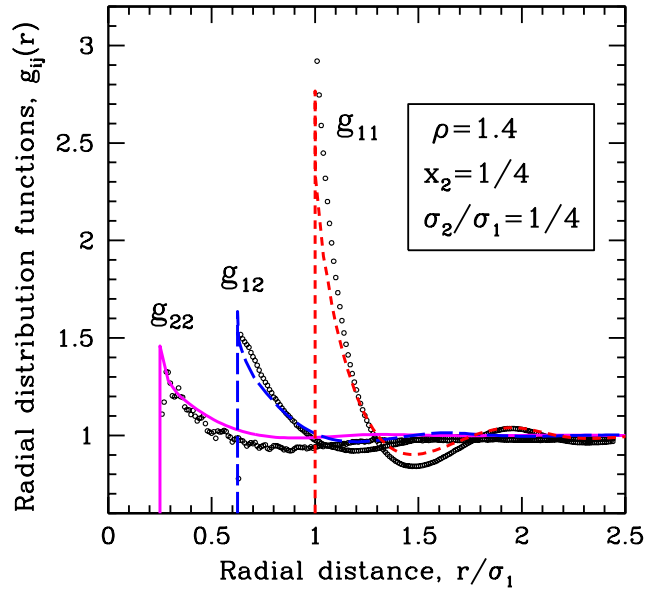}}
\caption{\label{f.g_rho14} (Color online) Radial distribution functions for a binary mixture
in $d=5$ {dimensions} with parameters $x_2=\frac{1}4$, $\sigma_2/\sigma_1=\frac{1}4$ and density
$\rho \sigma_1^5=1.4$ ($\eta=0.172774$). Solid lines: results from the {RFA-PY} method.
{Symbols: computer simulations \protect\cite{GAH01}.}}
\end{figure}
\begin{figure}
\scalebox{0.42}{\includegraphics{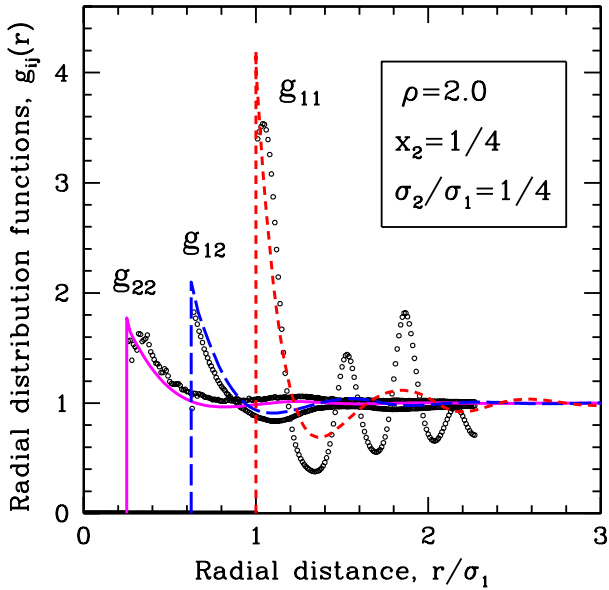}}
\caption{\label{f.g_rho20}
(Color online) As in Fig.\ \protect\ref{f.g_rho14} but for $\rho \sigma_1^5=2.0$ ($\eta=0.246820$).}
\end{figure}

Figures \ref{f.g_rho14} and \ref{f.g_rho20} show the radial distribution
functions $g_{ij}(r)$ of a highly asymmetric binary 5D-sphere mixture with
$\sigma_2/\sigma_1=\frac{1}{4}$ and $x_2=\frac{1}{4}$, at the reduced densities
$\rho \sigma_1^5=1.4$ and $\rho\sigma_1^5=2.0$, respectively.
The selected densities lie in the proximity of the phase transition for
the pure one-component {fluid}
($\eta_f=0.19$) predicted by Michels and Trappeniers \cite{MT84}.
{We see that the RFA-PY solution} at $\eta\approx 0.17$ gives
reasonable good values for all pair functions throughout the $r$-region
(see Fig.\ \ref{f.g_rho14}). At a higher packing fraction, $\eta\approx 0.25$,
we observe in Fig.\ \ref{f.g_rho20} that similar quantitative trends
in $g_{22}$ and $g_{12}$ are followed by both simulations and {RFA-PY}
solutions. {However}, there are discrepancies particularly severe in
the values of $g_{11}$ corresponding to pairs of  big particles.
The strong oscillations observed in the simulation data of $g_{11}$
can be considered as a signature of a solid phase at this density
\cite{GAH01}, which is not captured by the {RFA-PY} solution.

\section{Conclusions} \label{s.conclus}

In this paper we have extended the {RFA} method
to multicomponent systems of hard-hyperspheres in odd dimensions.
{The main features} of this approach are based on simple physical considerations
on Laplace functionals $G_{ij}(s)$ of the radial distribution functions,
which are closely related to the structure functions of the fluid.
The basic {physical requirements upon which the method is based are}:
(i) The radial distribution functions take finite values at contact
and vanish {inside the core}, which implies that
$\lim_{s\rightarrow \infty} [s^{(5-d)/2}
e^{\sigma_{ij} s} G_{ij}(s)]=\text{finite}$;
(ii) the isothermal compressibility is finite, which implies
$\lim_{s\rightarrow 0} [G_{ij}(s)-(d-2)!!/s^2]=\text{finite}$; and
(iii) {the first three terms in the series expansion in powers of
density of the structure factors are exact}.
Condition (iii) involves the evaluation of the overlap volume of
two arbitrary $d$-spheres. For space of odd dimensionality, we have found
the  exact, analytical and closed-form
expression  for the overlap volume of two hyperspheres of arbitrary sizes as
a function of the center-distance [cf.\ Eqs.\ (\ref{e.Omega}) and (\ref{e.R4n4g})].
We {have been able to perform this evaluation thanks to} the use of reverse Bessel polynomials
and Fourier analysis.

The primary result of the present work {has been} to provide a theoretical method
for the evaluation of thermodynamic and structural quantities of
multicomponent mixtures of hard hyperspheres at odd dimensions. We have shown
that this approach gives the exact solution of the OZ equation
with the PY closure. {From that perspective, our work extends, on  one hand,
the Lebowitz solution for 3D-sphere mixtures \cite{L64}  to higher dimensions and, on the other hand, the PY solution for one-component hyperspheres \cite{FI81,L84,RHS04,RHS07} to the case of mixtures.}

Although the theory here developed is general and applies to any
odd-dimensional, hard-hypersphere fluid with an arbitrary number of components,
the results in this paper concentrate on 5D binary mixtures.
{We have checked that the virial ($Z_v$) and compressibility ($Z_c$) routes to the compressibility factor under- and overestimate, respectively, the simulation data, the  interpolation approximation $Z=\frac{2}{5}Z_v+\frac{3}{5}Z_c$ providing excellent results.
In addition, reasonable agreement between RFA-PY results and simulation data for the radial distribution functions are found}
at densities lower than {that of the fluid-solid} phase transition predicted for a pure fluid
at this dimension.

{The implementation of the RFA method developed here involves as many unknowns as the minimum number required by the physical conditions and that is why it coincides with the PY solution. However, as done for 3D mixtures \cite{YSH98} and for $d$-dimensional one-component systems \cite{RS07}, one can go beyond the PY level by adding an extra matricial term $\sm{L}_{n+2}$ in Eqs.\ \eqref{Lserie} and \eqref{Bdef}, and replacing $\sm{I}$ by $(1+u s)\sm{I}$ in Eq.\ \eqref{Bdef}. The elements of  $\sm{L}_{n+2}$ and the parameter $u$ are free and can be fixed, for instance, by imposing given expressions for the contact values $g_{ij}(\sigma_{ij}^+)$  and the thermodynamically consistent isothermal susceptibility
 $\chi$.}

{We expect that the results presented in this paper can stimulate simulation studies on multicomponent systems of hard hyperspheres. We also plan to undertake the investigation of possible fluid-fluid demixing transitions predicted by the equations of state obtained here within the RFA-PY approach.}

\acknowledgments
We would like to thank M. Gonz\'alez-Melchor, {who} kindly sent us her
simulation data which {were} used in Figs.\ \ref{f.g_rho14} and \ref{f.g_rho20}.
{The work of {R.D.R.} has been supported by the Consejo Nacional de Investigaciones Cient\'ificas y
T\'ecnicas (CONICET, Argentina) through Grant
PIP 112-200801-01474.  A.S. acknowledges support from the Ministerio de Ciencia e Innovaci\'on (Spain) through Grant No. FIS2010-16587, partially financed by FEDER (Fondo Europeo de Desarrollo Regional) funds.}

\appendix
\section{Fourier Transform} \label{a.fourier}

In odd-dimension spaces, the Fourier transform of an arbitrary radial function $\xi(r)$
and the inverse operation can be evaluated in terms of the reverse Bessel
polynomial $\theta_n(t)$ of degree $n=(d-3)/2$ as \cite{RS07}
\beq \label{e.Fou1}
\widehat{\xi}(k)=\frac{(2\pi)^{(d-1)/2}}{k^{d-2}} \ii \int_{-\infty}^\infty \dd r\,
r \xi(r) \theta_n(\ii kr) e^{-\ii kr},
\eeq
\beq \label{e.Fou1b}
\xi(r)=\frac{(2\pi)^{-(d+1)/2}}{r^{d-2}} \ii \int_{-\infty}^\infty \dd k\,
k \widehat{\xi}(k) \theta_n(\ii kr) e^{-\ii kr}.
\eeq
Alternatively,
\beq \label{e.Fou2}
\widehat{\xi}(k)=\frac{(2\pi)^{(d-1)/2}}{k^{d-2}} 2\Im\{\cm{F}_n[\xi(r)](-\ii k)\},
\eeq
\beq \label{e.Fou2b}
\xi(r)=\frac{(2\pi)^{-(d+1)/2}}{r^{d-2}} 2\Im\{\cm{F}_n[\widehat{\xi}(k)](-\ii r)\},
\eeq
where $\cm{F}_n$ is a Laplace functional defined by
\beq
\cm{F}_n[\xi(r)](s)\equiv \int_0^\infty \dd x\, x\xi(x)\theta_n(sx)e^{-sx}.
\eeq
From an integral relation on $\theta_n(t)$ one arrives to \cite{RS07}
\beq \label{e.F1}
\cm{F}_n[1](s)= \frac {\theta_{n+1}(0)}{s^2}=\frac{(2n+1)!!}{s^2},
\eeq
\beq \label{e.F2}
\cm{F}_n[\Theta(r-a)](s)= \frac {\theta_{n+1}(as) e^{-as}}{s^2}.
\eeq
%

\section{Co-volume of two hyperspheres} \label{a.covolume}

The overlap volume of two hyperspheres of radii $a$ and $b$, whose centers
are a distance $r=|\bm{r}|$, can be evaluated as
\beq
\Omega_{a,b}(r)=\int \dd\bm{r}' \Theta(a-|\bm{r}'|)\Theta(b-|\bm{r}'-\bm{r}|).
\label{B1}
\eeq
Henceforth, without loss of generality, we assume $a\le b$. The right-hand side of Eq.\ \eqref{B1} is a convolution. Thus, in  Fourier space Eq.\ \eqref{B1}
reads
\beq
\widehat{\Omega}_{a,b}(k)=\widehat{\Theta}_a(k)\widehat{\Theta}_b(k),
\label{B2}
\eeq
where we have called
\beq
\widehat{\Theta}_a(k)\equiv \int \dd\mathbf{r}\,\Theta(a-r)e^{-\ii \mathbf{k}\cdot\mathbf{r}}.
\label{B3}
\eeq
Now  using Eqs.\ \eqref{e.Fou2}, (\ref{e.F1}), and (\ref{e.F2}), and taking into account that $\Theta(a-r)=1-\Theta(r-a)$, we get
\beqn
\widehat{\Theta}_a(k)&=&\frac{(2\pi)^{(d-1)/2}}{k^{d-2}}
2\Im\left[ \frac{(2n+1)!!}{s^2} \right.\nn
&&\left.-\frac{\theta_{n+1}(a s) e^{-a s}}
{s^2} \right]_{s=-\ii k} \nn
&=&\frac{(2\pi)^{(d-1)/2}}{k^{d}}\ii \left[
\theta_{n+1}(-\ii ka) e^{\ii ka}\right.\nn
&&\left.-\theta_{n+1}(\ii ka) e^{-\ii ka}
\right] .
\eeqn
Therefore,
\beqn
\widehat{\Omega}_{a,b}(k) &=& \frac{(2\pi)^{d-1}}{k^{2d}} \left[
    \theta_{n+1}(\ii ka) \theta_{n+1}(-\ii kb) e^{\ii k(b-a)}   \right. \cr &&
   +\theta_{n+1}(-\ii ka) \theta_{n+1}(\ii kb) e^{-\ii k(b-a)}  \cr &&
   -\theta_{n+1}(-\ii ka) \theta_{n+1}(-\ii kb) e^{\ii k(b+a)}  \cr && \left.
   -\theta_{n+1}(\ii ka) \theta_{n+1}(\ii kb) e^{-\ii k(b+a)} \right].
\eeqn
We can obtain $\Omega_{a,b}(r)$ using the inverse Fourier transform
(\ref{e.Fou1b}). Integration on the complex plane and the application of
the residue theorem yield
\beqn \label{e.Omega0}
\Omega_{a,b}(r)&=&\frac{(2\pi)^{(d-1)/2}}{r^{d-2}} \left[
R^{(a,b)}_{4n+4}(r) \Theta(b+a-r) \right.\cr
&&\left. -R^{(-a,b)}_{4n+4}(r) \Theta(b-a-r) \right],
\eeqn
with
\beqa
 \label{e.R4n4}
R^{(\pm a,b)}_{4n+4}(r) &=& \mathop{\mbox{Res}}_{t=0}
\left[\frac  {\theta_{n+1}(\mp ta)
\theta_{n+1}(-tb) \theta_{n}(tr) e^{t(b\pm a-r)}} {t^{4n+5}} \right]\nn
&=&\frac{1}{(4n+4)!}\left[\partial_t^{4n+4}\theta_{n+1}(\mp ta)
\theta_{n+1}(-tb) \right.\nn
&&\left.\times\theta_{n}(tr) e^{t(b\pm a-r)}\right]_{t=0}.
\eeqa
Equation \eqref{e.R4n4} implies that  $R^{(\pm a,b)}_{4n+4}(r)$ are polynomials of degree $4n+4=2d-2$.
If $r<b-a$ the smaller hypersphere will be fully contained within the bigger one,
so that $\Omega_{a,b}(r)$ will be equivalent to the volume of the former, i.e.,
\beq
\Omega_{a,b}(r)= (2a)^d v_d = \frac{2a^d (2\pi)^{(d-1)/2}}{d!!},\quad r\le b-a.
\eeq
Thus, according to (\ref{e.Omega0}), the difference between the polynomials
$R_{4n+4}^{(\pm a,b)}$ takes the following expression (for any $r$)
\beq
R_{4n+4}^{(a,b)}(r)-R_{4n+4}^{(-a,b)}(r)= \frac{2a^{2n+3} r^{2n+1}}{(2n+3)!!}.
\eeq
Therefore, the final result is
\beq \label{e.Omega}
\Omega_{a,b}(r) = \begin{cases}
\frac{(2\pi)^{(d-1)/2}}{d!!} 2a^d, & r \le b-a, \\
(2\pi)^{(d-1)/2}\frac{R_{4n+4}^{(a,b)}(r)}{r^{d-2}},   & b-a \le r \le b+a,  \\
0,  &r \ge b+a.
\end{cases}
\eeq

Let us now obtain an explicit expression for the polynomial $R_{4n+4}^{(a,b)}(r)$. According to Eq.\ \eqref{e.R4n4}, it is the coefficient of $t^{4n+4}$ in the expansion of
$\theta_{n+1}(-ta)\theta_{n+1}(-tb)\theta_{n}(tr) e^{t(b+a-r)}$ in powers of $t$.
First,  note that
\beq
\theta_{n+1}(-ta) \theta_{n+1}(-tb)
= \sum_{\ell=0}^{2n+2} c_{n,\ell}^{(a,b)} t^\ell,
\eeq
where
\beq
c_{n,\ell}^{(a,b)} = (-1)^\ell\sum_{\ell_1=\max(0,\ell-n-1)}^{\min(\ell,n+1)} \omega_{n+1,\ell_1}\omega_{n+1,\ell-\ell_1}
a^{\ell_1} b^{\ell-\ell_1}.
\eeq
Next,
\beq
\theta_{n+1}(-ta) \theta_{n+1}(-tb) \theta_n(tr)
= \sum_{\ell=0}^{3n+2} d_{n,\ell}^{(a,b)}(r) t^\ell,
\eeq
with
\beq
d_{n,\ell}^{(a,b)}(r)=\sum_{\ell_1=\max{(0,\ell-2n-2)}}^{\min(\ell,n)}\omega_{n,\ell_1}c_{n,\ell-\ell_1}^{(a,b)}r^{\ell_1}.
\label{B4}
\eeq
Finally, the coefficient of $t^\ell$ in the $t$-expansion of  $\theta_{n+1}(-ta)\theta_{n+1}(-tb)\theta_{n}(tr) e^{t(b+a-r)}$ is
\beq
\sum_{\ell_1=\max(0,\ell-3n-2)}^\ell \frac{(a+b-r)^{\ell_1}}{\ell_1!} d_{n,\ell-\ell_1}^{(a,b)}(r).
\eeq
Setting $\ell=4n+4$ and inserting Eq.\ \eqref{B4}, one obtains
\beqa
\label{e.R4n4g}
R_{4n+4}^{(a,b)}(r)&=&(a+b-r)^{n+2}\sum_{\ell=0}^{3n+2}\frac{(a+b-r)^{3n+2-\ell}}
{(4n+4-\ell)!} \cr &&\times
\sum_{\ell_1=\max(0,\ell-2n-2)}^{\min(\ell,n)}\omega_{n,\ell_1}c_{n,\ell-\ell_1}^{(a,b)} r^{\ell_1}.
\eeqa
Thus, for example, at $d=3,5$ (respectively $n=0,1$) one finds
\beq
\label{R4}
R_{4}^{(a,b)}(r) = \frac{(a+b-r)^2}{24} \left[r^2+2(a+b)r-3(b-a)^2\right],
\eeq
\beqa
\label{R8}
R_{8}^{(a,b)}(r) &=& \frac{(a+b-r)^3}{1920} \left[3r^5 +9(a+b)r^4
\right. \cr &&-2(a^2-18ab+b^2)r^3
-2(b-a)^2(a+b)r^2 \nn
 &&\left. +15(b-a)^4 r+5(b-a)^4(a+b)\right].
\eeqa

It can be checked that $R_{4n+4}^{(a,b)}(r)$ admits the following structure:
\beqa
R_{4n+4}^{(a,b)}(r)&=&(a+b-r)^{n+2}\left[r^{2n+1}P_{n+1}^{(a,b)}(r)\right.\nn
&&\left.+(b-a)^2K_{2n}^{(a,b)}(r)\right],
\eeqa
where $P_{n+1}^{(a,b)}(r)$ and $K_{2n}^{(a,b)}(r)$ are polynomials of degree $n+1$ and $2n$, respectively.
Clearly, in the case of identical hyperspheres ($a=b$)
$R_{4n+4}^{(a,a)}(r)$ adopts the known expression
\beq
R_{4n+4}^{(a,a)}(r)=(2a-r)^{n+2} r^{2n+1}P_{n+1}^{(a,a)}(r),
\eeq
with $P_{n+1}^{(1,1)}(r)$ given by Eqs.\ (B7) and (B8) of Ref.\ \cite{RS07}.

\section{Properties of $\sm{B}(s)$} \label{s.Bs}

The series expansion of the function $\phi(x)$ defined by Eq.\ \eqref{phi} is
\beq \label{phi_s1}
\phi_m(x) = -\sum_{\ell=1}^\infty \frac{(-1)^{m+\ell}}{(m+\ell)!} x^\ell.
\eeq
Therefore, Eq.\ (\ref{phimatrix}) yields
\beq \label{phi_s2}
\sm{\Phi}_m(s) = \sum_{\ell=1-\delta}^\infty \sm{C}_{d+\ell-m} s^\ell\quad (\delta=0,1),
\eeq
where the diagonal matrices $\sm{C}_m$ are
\beq \label{mez_C}
\left(\sm{C}_m\right)_{ii}= -(-2\pi)^{(d-1)/2} x_i\,\frac{(-\sigma_i)^{m}}{m!}.
\eeq
Thus, the series expansion of the matrix $\sm{B}(s)$ defined by Eq.\ (\ref{Bdef}) is
\beq \label{BBk}
\sm{B}(s)=\sum_{\ell=0}^\infty \sm{B}_\ell \,s^\ell
\eeq
with
\beq \label{B0}
\sm{B}_0 =
\begin{cases} \sm{I} & (\delta=0), \\
 \sm{I}+\rho\sum_{m=0}^{n+1} \sm{C}_{d-m}\cdot\sm{L}_m & (\delta=1),
\end{cases}
\eeq
\beq \label{Bk}
\sm{B}_{\ell}=\rho \sum_{m=0}^{n+1} \sm{C}_{d+\ell-m}\cdot\sm{L}_m \quad (\ell\ge 1).
\eeq
According to Eqs.\ \eqref{phi_s2} and \eqref{BBk},
\beq
\lim_{s\rightarrow 0} \sm{\Phi}_m(s)=
\begin{cases}
0 & (\delta=0), \\
 \sm{C}_{d-m} & (\delta=1),
\end{cases}
\eeq
\beq
\lim_{s\rightarrow 0} \sm{B}(s)=
\begin{cases} \sm{I} & (\delta=0), \\
 \sm{I}+\rho\sum_{m=0}^{n+1} \sm{C}_{d-m}\cdot\sm{L}_m & (\delta=1).
 \end{cases}
 \label{Bs0}
\eeq

As for the behaviors in the limit $s\to\infty$, it is straightforward from Eqs.\ \eqref{Bdef}--\eqref{phi} to get
\beq
\lim_{x\to\infty}\phi_m(x)=\frac{(-1)^m}{m!},
\eeq
\beq
\lim_{s\to \infty} \sm{\Phi}_m(s)=
\begin{cases}
-\sm{C}_{d-m} & (\delta=0), \\
 0 & (\delta=1),
\end{cases}
\eeq
\beq
\lim_{s\to \infty} \sm{B}(s)=
\begin{cases}
\sm{I}-\rho\sum_{m=0}^{n+1} \sm{C}_{d-m}\cdot\sm{L}_m & (\delta=0), \\
 \sm{I} & (\delta=1).
\end{cases}
\label{Bsinfty}
\eeq

\section{The Rational Function Approximation in the low-density limit}
\label{s.lowdensity}

Let us consider the low-density expansions of  $\sm{L}_m$ and $\sm{B}(s)$,
\beq
\sm{L}_m=\sm{L}_m^{(0)}+\rho \sm{L}_m^{(1)}+O(\rho^2),
\label{7}
\eeq
\beq
\sm{B}(s)=\sm{I}+\rho\sm{B}^{(1)}(s)+O(\rho^2),
\label{8}
\eeq
where
\beq
{B}_{ij}^{(1)}(s)=\sum_{m=0}^{n+1}\left[\sm{\Phi}_m(s)\right]_{ii}\left[\sm{L}_m^{(0)}\right]_{ij}.
\label{9}
\eeq
Insertion into Eq.\ \eqref{Grfa} yields Eq.\ \eqref{e.G01}  with
\beq
G_{ij}^{(0)}(s)=\frac{e^{-\sigma_{ij}s}}{s^2}\sum_{m=0}^{n+1}\left[\sm{L}_m^{(0)}\right]_{ij} s^m,
\label{10}
\eeq
\beq
G_{ij}^{(1)}(s)=\frac{e^{-\sigma_{ij}s}}{s^2}\sum_{m=0}^{n+1}\left[\sm{L}_m^{(1)} -\sm{L}_m^{(0)}\cdot\sm{B}^{(1)}(s)\right]_{ij} s^m.
\label{11}
\eeq
Comparison between Eqs.\ \eqref{e.G0} and Eq.\ \eqref{10} implies that
\beq
\left[\sm{L}_m^{(0)}\right]_{ij}=\omega_{n+1,m}\sigma_{ij}^m.
\label{D1}
\eeq

Let us consider now the matrices $\sm{L}_m^{(1)}$. It is convenient to express their elements as
\beq
\left[\sm{L}_m^{(1)}\right]_{ij}=\sum_{\ell=1}^\NN x_\ell L_{m,ij\ell}^{(1)}.
\eeq
Using Eqs.\ \eqref{phimatrix} and \eqref{phi}, one can recast Eq.\ \eqref{11}  into the form \eqref{e.G1} with
\beqa
G_{ij\ell}^{(1)}(s)&=&\frac{e^{-\sigma_{ij}s}}{s^2}\sum_{m=0}^{n+1}L_{m,ij\ell}^{(1)} s^m+\frac{\nu_d}{s^{d-2}} G_{i\ell}^{(0)}(s)G_{j\ell}^{(0)}(s)
\nn &&
-
\nu_d\frac{e^{-\sigma_{ij}s}}{s^{d+2}}\theta_{n+1}(\sigma_{i\ell} s)\sum_{m=0}^{n+1}\omega_{n+1,m}\left(\sigma_{j\ell} s\right)^m
\nn
&&\times\sum_{q=0}^{d-m-\delta}\frac{(-\sigma_\ell s)^q}{q!}.
\label{D2}
\eeqa
Comparison with Eqs.\ \eqref{6} and \eqref{3.1} allows us to identify
\beq
\sum_{m=0}^{n+1}L_{m,ij\ell}^{(1)} s^m=\nu_d s \overline{Q}_{ij\ell}(s)\quad (\delta=0),
\label{D3}
\eeq
\beqa
\sum_{m=0}^{n+1}L_{m,ij\ell}^{(1)} s^m&=&\nu_d s \overline{Q}_{ij\ell}(s)-\nu_d\theta_{n+1}(\sigma_{i\ell}s)\sum_{m=0}^{n+1}\frac{\omega_{n+1,m}}{(d-m)!}\nn
&&\times
\sigma_{j\ell}^m(-\sigma_\ell)^{d-m}\quad (\delta=1).
\label{D4}
\eeqa

Therefore, the RFA proposal \eqref{Grfa}--\eqref{phi} is consistent with the exact result to first order in density.

\section{Pure-fluid limit} \label{s.one}

We consider here $\sigma_{ij}=\sigma=1$ $\forall$ $i,j$, within the case
$\delta=0$. Then $\rho\sm{\Phi}_m(s) = \lambda_d \phi_{d-m}(s) \lx \eta_k \rx$ and
one can expect that $\sm{L}_m=a_m \lx 1 \rx$, with $a_m$ being a certain
constant. {Here we  use the matrix notation defined in Eq.\ \eqref{notation} and have defined
\beq
\lambda_d\equiv\frac{\nu_d}{v_d}=(-1)^{(d-1)/2}2^{d-1}d!!.
\eeq
}
Thus,
\beq \label{L_one}
\sm{L}(s)= \sum_m a_m s^m \lx 1 \rx
\eeq
and
\beq
\sm{B}(s)=\sm{I} +\lambda_d\sum_{m=0}^{n+1}  \phi_{d-m}(s) a_m \lx \eta_i \rx .
\eeq
Using the {mathematical property (\ref{invA}) and taking into account that} $\text{tr}(\lx \eta_i \rx)=\eta$,
\beq \label{Binv_one}
\sm{B}^{-1}(s) = \sm{I} -\lambda_d\frac {\sum_{m=0}^{n+1}  \phi_{d-m}(s) a_m}
{1+\lambda_d \eta \sum_{m=0}^{n+1} \phi_{d-m}(s) a_m } \lx \eta_i \rx.
\eeq
From Eqs.\ (\ref{L_one}) and (\ref{Binv_one}) it readily follows:
\beq
\sm{L}(s)\cdot\sm{B}^{-1}(s) =
\frac {\sum_{m=0}^{n+1} a_m s^m} {1+\lambda_d \eta \sum_{m=0}^{n+1} \phi_{d-m}(s) a_m }  \lx 1 \rx .
\eeq
{Finally, Eq.\ \eqref{Grfa} becomes}
\beq
G_{ij}(s)= \frac{e^{-s}}{s^2} \frac {\sum_m a_m s^m}
{1+\lambda_d \eta \sum_{m=0}^{n+1} \phi_{d-m}(s) a_m } \quad \forall i,j.
\eeq
{This coincides with the PY solution of the one-component fluid in $d=2n+3$ dimensions, as derived from the RFA in Ref.\  \cite{RS07}.}

\section{Inversion of matrices} \label{s.matrix}

{Let us consider square matrices  defined {following the notation of} Eq.\ \eqref{notation}. Then,  \aa{as proved below}, one has
\beq \label{invA}
\left(\sm{I}+\lx A_j\rx\right)^{-1}= \sm{I}-\frac {\lx A_j\rx}{1+a}, \quad
\left(\sm{I}+\lx A_i\rx\right)^{-1}= \sm{I}-\frac {\lx A_i\rx}{1+a},
\eeq
\begin{subequations}
\label{invCB}
\beq \label{invCB1}
\left(\sm{I}+\lx C_i B_j\rx\right)^{-1}= \sm{I}-\frac {\lx C_i B_j\rx}{1+\beta},
\eeq
\beq \label{invCB2}
\left(\sm{I}+\lx B_i C_j\rx\right)^{-1}= \sm{I}-\frac {\lx B_i C_j\rx}{1+\beta}.
\eeq
\end{subequations}
}
{
\beqa \label{invACB}
\left(\sm{I}+\lx A_j\rx+\lx C_i B_j\rx\right)^{-1}&=&
\sm{I}+\frac{1}{\gamma}\left\{ \alpha \lx B_j\rx -(1+a)\lx C_i B_j\rx\right.\nn
&&\left.-(1+\beta)\lx A_j\rx
+b\lx C_i A_j\rx \right\},\nn
\eeqa
\beqa \label{invABC}
\left(\sm{I}+\lx A_i\rx+\lx B_i C_j\rx\right)^{-1}&=&
\sm{I}+\frac{1}{\gamma}\left\{ \alpha \lx B_i\rx -(1+a)\lx B_i C_j\rx\right.\nn
&&\left.-(1+\beta)\lx A_i\rx
+b\lx A_i C_j\rx \right\},\nn
\eeqa
where
\beq \label{def_a}
a \equiv \text{tr}(\lx A_j\rx) = \text{tr}(\lx A_i\rx),\quad
\alpha \equiv \text{tr}(\lx C_iA_j\rx) = \text{tr}(\lx A_iC_j\rx),
\eeq
\beq \label{def_b}
b \equiv \text{tr}(\lx B_j\rx) = \text{tr}(\lx B_i\rx),\quad
\beta \equiv \text{tr}(\lx C_iB_j\rx) = \text{tr}(\lx B_iC_j\rx),
\eeq
\beq
\gamma\equiv (1+a)(1+\beta)-\alpha b.
\eeq
}

{Equations (\ref{invA}) and (\ref{invCB}) follow immediately from the properties
$\lx A_j\rx\cdot\lx A_j\rx= a\lx A_j\rx$, $\lx A_i\rx\cdot\lx A_i\rx= a\lx A_i\rx$,
$\lx B_i C_j\rx\cdot\lx B_i C_j\rx= \beta \lx B_i C_j\rx$, and
$\lx C_i B_j\rx\cdot\lx C_i B_j\rx= \beta \lx C_i B_j\rx$.}

{The proof of Eqs.\ (\ref{invACB}) and (\ref{invABC}) is somewhat more involved. We first multiplicate  $(\sm{I}+\lx A_j\rx)^{-1}$
 by $\sm{I}+\lx A_j\rx+\lx C_i B_j\rx$
[see Eq.\ (\ref{invA})],
\beqn  \label{teo2b}
(\sm{I}+\lx A_j\rx)^{-1}\cdot\left(\sm{I}+\lx A_j\rx+\lx C_i B_j\rx\right)
&=& \sm{I}+\lx C_i B_j\rx\nn
&&-\frac {\lx A_j\rx\cdot\lx C_i B_j\rx}{1+a}  \cr
&=& \sm{I}+\lx C_i' B_j\rx,
\eeqn
where in the last step we have used that $\lx A_j\rx\cdot\lx C_i B_j\rx= \alpha \lx B_j\rx$ and  have introduced
\beq
C_i'\equiv C_i-\frac{\alpha}{1+a}.
\label{x3}
\eeq
Thus, the left inverse of $\sm{I}+\lx A_j\rx+\lx C_i B_j\rx$ is
\beq
\left(\sm{I}+\lx C_i' B_j\rx\right)^{-1}\cdot (\sm{I}+\lx A_j\rx)^{-1}=\left(\sm{I}-\frac {\lx C_i' B_j\rx}{1+\beta'}\right)
\cdot\left(\sm{I}-\frac {\lx A_j\rx}{1+a}\right),
\label{x2}
\eeq
where
\beq \label{betap}
\beta'\equiv \text{tr}(\lx C_i' B_j\rx) = \beta -\frac{\alpha b}{1+a}
\eeq
and
use has been made of Eqs.\ \eqref{invA} and \eqref{invCB1}.
Taking  the property $\lx C_i' B_j\rx\cdot \lx A_j\rx=b\lx C_i' A_j\rx$ into account,  Eq.\ \eqref{x2} becomes
\beqa
\left(\sm{I}+\lx C_i' B_j\rx\right)^{-1}\cdot (\sm{I}+\lx A_j\rx)^{-1}&=&\sm{I}-\frac {\lx A_j\rx}{1+a}-\frac {\lx C_i' B_j\rx}{1+\beta'}\nn
&&+\frac{b}{\gamma}\lx C_i' A_j\rx,
\label{x4}
\eeqa
where we have used $(1+\beta')(1+a)=\gamma$. Finally, from Eq.\ \eqref{x3} it is straightforward to see that the right-hand side of Eqs.\ \eqref{invACB} and \eqref{x4} coincide.
}

{An analogous method can be used to prove that the right-hand side of Eq.\ \eqref{invACB} is also the right inverse of   $\sm{I}+\lx A_j\rx+\lx C_i B_j\rx$. Multiplying the latter matrix by
 $(\sm{I}+\lx{A_j}\rx)^{-1}$
[see Eq.\ (\ref{invA})], one gets
\beq
\left(\sm{I}+\lx A_j\rx+\lx C_i B_j\rx\right)\cdot(\sm{I}+\lx{A_j}\rx)^{-1}
= \sm{I}+\lx C_i B_j'\rx ,
\eeq
with
\beq
{B}_j'\equiv{B_j}-\frac{b}{1+a}A_j.
\label{x6}
\eeq
Thus, the right inverse of $\sm{I}+\lx A_j\rx+\lx C_i B_j\rx$ is
\beqa
(\sm{I}+\lx{A_j}\rx)^{-1}\cdot\left(\sm{I}+\lx C_i B_j'\rx\right)^{-1}&=&\sm{I}-\frac {\lx A_j\rx}{1+a}-\frac {\lx C_i B_j'\rx}{1+\beta'}\nn
&&+\frac{\alpha}{\gamma}\lx B_j'\rx.
\label{x5}
\eeqa
Here we have used $\beta'=\text{tr}(\lx C_i B_j'\rx$ and $\lx A_j\rx\cdot\lx C_i B_j'\rx=\alpha\lx B_j'\rx$. Replacement of Eq.\ \eqref{x6} makes the right-hand side of Eq.\ \eqref{x5} coincide with the right-hand side of Eq.\ \eqref{invACB}. This completes the proof of Eq.\ \eqref{invACB}. Equation \eqref{invABC} is just the transpose of Eq.\ \eqref{invACB}. Note finally that Eq.\ \eqref{invACB} reduces to Eqs.\ \eqref{invA} and \eqref{invCB1} by particularizing to $C_i=0$ and $A_i=0$, respectively.}



\begin{thebibliography}{99}
\bibitem{note}
Note that if $\sigma_{ij}=1$ $\forall i,j$, then Eq.\ (\ref{d5_L1})
yields the relation $a_1=3[1-\eta(1+10a_2)]/(1-6\eta)$
($\sm{L}_\ell=a_\ell\lx 1\rx$ $\ell=1,2$) earlier obtained for one-component fluids
\protect\cite{RS07}.
\end{thebibliography}
\end{document}